%% file: varga_IAMPDM_fin_2026_arxiv.tex
\documentclass[final,5p,times]{elsarticle}
\biboptions{numbers,square,sort&compress}

\usepackage{amssymb}
\usepackage{amsmath}
\usepackage{siunitx}

\usepackage{pgfplots}
\pgfplotsset{compat=newest}

\usetikzlibrary{
	pgfplots.statistics,
}

\usetikzlibrary{plotmarks}
\usetikzlibrary{arrows.meta}
\usepgfplotslibrary{patchplots}
\usepackage{grffile}
\usepackage{tikz}
\usepackage{tikzscale}
\usepackage{placeins}

\usepackage{subcaption} 
\usepackage{graphicx} 
\usepackage{caption}    
\usepackage{subcaption}  
\usepackage{siunitx}     
\usepackage{float}      

\usepackage{xcolor} 
\usepackage[colorlinks=true, allcolors=blue]{hyperref}
\usepackage{todonotes}

\usepackage{algorithmic}
\usepackage[linesnumbered,ruled,vlined]{algorithm2e}

\newcommand{\vek}[1]{{\mathbf #1}}

\newcommand{\sv}[1]{\boldsymbol{#1}}

\newcommand{\ml}[1]{\mathrm{#1}} 
\newcommand{\cc}[1]{\cite{#1}}

\journal{Elsevier}

\begin{document}
	
	\begin{frontmatter}
		
		\title{Interaction-Aware Model Predictive Decision-Making for Socially-Compliant Autonomous Driving in Mixed Urban Traffic Scenarios} 
		
		\author[label0]{Balint Varga} 
		\affiliation[label0]{
			addressline={balint.varga2@kit.edu},
			organization={Institute of Control Systems, Karlsruhe Institute for Technology},
			city={Karlsruhe},
			postcode={D-76131}, 
			country={Germany}}
		
		\author[label1]{Thomas Brand}
		\author[label1]{Marcus Schmitz}
		\affiliation[label1]{
			organization={Wuerzburg Institute for Traffic Sciences GmbH},
			city={Veitshoechheim},
			postcode={D-97209},
			country={Germany}}
		
		\author[label2]{Ehsan Hashemi}
		\affiliation[label2]{
			organization={Faculty of Engineering - Mechanical Engineering Dept, University of Alberta},
			city={Edmonton AB},
			postcode={Ca-T6G 2H5},
			country={Canada}}

		\begin{abstract}
			Autonomous vehicles must negotiate with pedestrians in ways that are not only safe, but also legible and socially acceptable. This paper presents an interaction-aware model predictive decision-making (IAMPDM) framework that couples a gap-acceptance-inspired pedestrian intention model with model predictive control (MPC) to coordinate intent reasoning and vehicle control in real time. The pedestrian model maps time-to-collision (TTC) to a continuous crossing-likelihood and incorporates the discounting of the pedestrian's crossing intention to represent hesitation and potential standstills. The resulting signals then scale MPC safety costs and minimum-distance constraints. We implement IAMPDM in a projection-based, motion-tracked simulator and compare it with a rule-based intention-aware controller (RBDM) and a conservative non-interactive baseline (NIA). In a human-in-the-decision-loop study with 25 participants (one excluded from subjective analyses), intention-aware methods generally reduced scenario completion time relative to NIA in the tested scenarios, while also producing tighter surrogate safety margins (e.g., TTC- and distance-based measures). Across most metrics, IAMPDM and RBDM were not statistically distinguishable, except for TTC in one scenario. Participants' subjective ratings tended to favor the intention-aware approaches over the non-interactive baseline. Finally, we translate these simulator observations into concrete tuning guidance.			
			
		\end{abstract}

		
	\end{frontmatter}
	
	\section{Introduction}
\begin{figure}[t!]
		\centering
		\includegraphics[width=0.99\linewidth]{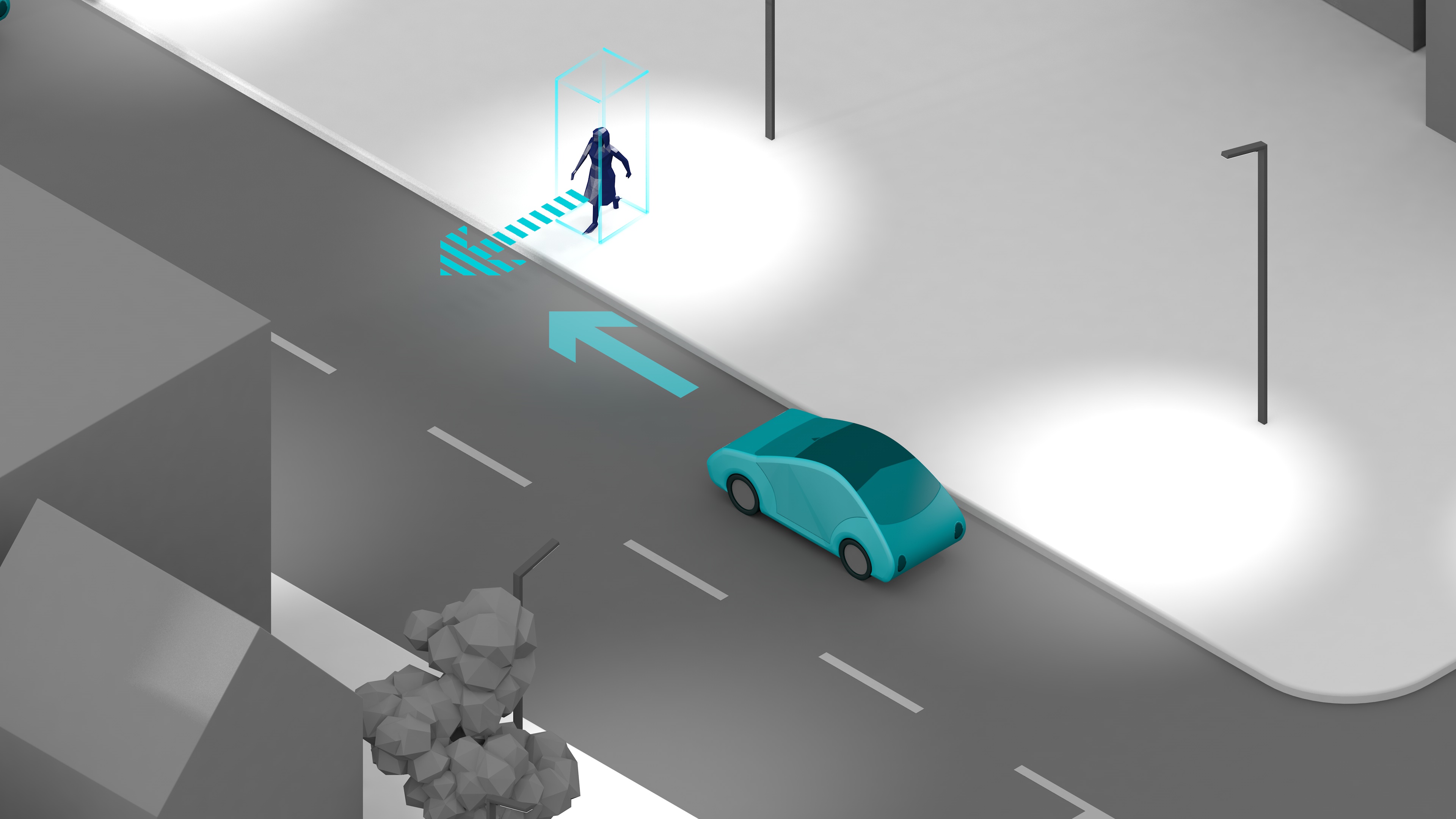}
		\caption{The pedestrian crosses the street at an unsignalized intersection interacting with an autonomous vehicle. With the courtesy of version1 GmbH}
		\label{fig:scenario_1}
	\end{figure}

	Highly automated and autonomous vehicles (AVs) are increasingly becoming a part of our daily lives \cc{2022_FactorsAffectingPedestrians_zhou}. 
	Integrating these systems into society largely depends on the trust of vulnerable road users, such as cyclists and pedestrians, whose safety is essential \cc{2023_IntAware_Merg_Moh}. 
	Incidents involving automated driving functions often make headlines, fostering public skepticism towards these technologies. 
	Consequently, significant research efforts are directed at enhancing automated vehicles with advanced communication channels and decision-making algorithms designed to manage complex interaction scenarios, see \cc{markkula2020defining}. For example, in urban environments and city centers -- where vehicles move at slower speeds, and pedestrians may cross unexpectedly~-- an effective human-machine interaction is critical to fostering trust in these systems. 
	Figure \ref{fig:scenario_1} illustrates a typical urban scenario where a pedestrian crosses the street at an unsignalized intersection while interacting with an autonomous vehicle.
	
	In recent years, advances in machine learning and control~-- especially game-theoretic models and reinforcement learning (RL)~-- have improved automated vehicles’ ability to interact with pedestrians and cyclists by capturing the negotiation and decision dynamics that arise in mixed-traffic situations, \cc{2018_WhereAreDangerous_hu,2020_PedestriansRoadCrossing_razmirad,2023_SharedSpaceMultimodal_kampitakis,2023_WhoGoesFirst_kalantari}
	The performance and safety of these interaction strategies reported in the literature are commonly evaluated under two principal paradigms. The first is pure simulation - large-scale, repeatable virtual environments that allow researchers to explore many scenarios and parameter settings efficiently \cc{2022_CautiousBehaviorsPedestrians_luu,2023_SharedSpaceMultimodal_kampitakis}. 
	The second paradigm embeds humans into the evaluation loop, either through human-in-the-loop studies or controlled user experiments, which can reveal important aspects of human behavior, trust, and perceived safety that simulations alone cannot capture \cc{2021_ReviewVirtualReality_tran,2021_InteractionPedestriansAutomated_bindschadel}.

	While both approaches are indispensable, literature-based experiments typically impose constraints on the decision space: vehicle actions (e.g., braking profiles, trajectory choices) and the range of participants’ responses are often predefined to make analysis tractable and results reproducible. This strong experimental control facilitates statistical comparison across algorithms and conditions but comes at the cost of external validity - the restricted set of behaviors and responses can fail to capture the unpredictability and heterogeneity of real-world interactions. As a result, findings derived from such controlled setups may overestimate robustness or underrepresent rare but safety-critical behaviors, limiting the applicability of conclusions when algorithms are deployed outside laboratory conditions.

	Therefore, in this work, we address those limitations by proposing and validating an interaction aware model predictive decision-making (IAMPDM) approach in a simulator using a true human-in-the-decision-loop setup where participants are free to choose their actions. 
	Simulator studies are a practical and ethically safer alternative to on-road trials for examining unconstrained human–AV negotiation \cc{2021_ReviewVirtualReality_tran}. 
	Building on this rationale, the present paper reports a simulator experiment that preserves participant freedom while enabling a systematic comparison of decision-making strategies. 
	The main contributions are:
	\begin{itemize}
		\item We adapt the IAMPDM for real-time experiments,
		\item We design a novel human-in-the-decision-loop simulator study that balances naturalistic validity and experimental control.
		\item Finally, we conduct an experiment with 25 participants, and analyze the findings to derive recommendations for real - world implementation of IAMPDM.
	\end{itemize}
	
	The paper is structured as follows: In Section \ref{sec:state_of_the_art}, the state of the art of intention-aware automated vehicles is presented.
	Section \ref{sec:mpc_algo} presents the concept of the IAMPDM. The simulator setup of the validation and the experimental design are given in Section \ref{sec:experiment}. Section \ref{sec:Results_discussion} provides the results of the experiment and a further discussion about the limitations. Finally, the paper is summarized in Section~\ref{sec:summary}.

	\section{State of the Art} \label{sec:state_of_the_art}
	The upcoming sections review key conceptual frameworks, contrasts model-free and model-based approaches, and discusses validation methods for autonomous vehicle decision-making algorithms, with emphasis on interaction-aware systems.
	
	\subsection{Conceptual Frameworks for Interaction}
	
	Defining “interaction” in the context of mixed human-\, automated traffic is a central challenge for the development of advanced driver-assistance and autonomous vehicle systems. 
	According to the conceptual framework presented by \cc{markkula2020defining}, interaction encompasses any situation in which the actions of one road user are shaped by the observed or anticipated behavior of another. This includes not only direct collision avoidance but also the exchange of information, either implicitly through kinematic cues (such as deceleration profiles or time-to-collision), or explicitly via signals and gestures. The framework of \cc{markkula2020defining} distinguishes between \textit{implicit communication} - for example, the yielding intent communicated by gradual deceleration - and \textit{explicit communication}, such as eye contact or hand gestures. The literature increasingly emphasizes the importance of both, as urban traffic negotiation is a continuous, communicative process involving prediction, negotiation, and mutual adaptation.
	
	Recent surveys, \cc{wang2022social} and \cc{crosato2024social}, catalogue the broad range of computational models developed for social interaction in AVs. These reviews cover approaches from game theory, which model strategic reasoning between agents, to frameworks grounded in social psychology, which seek to explain and simulate human-like negotiation, cooperation, and norm adherence in traffic.
	
	\subsection{Decision-Making Concepts}
	
	\subsubsection{Model-Free, Learning-Based Approaches}
	
	Model-free approaches do not rely on physics-based or \linebreak logic-based models. Instead, they learn optimal behaviors \linebreak through trial and error by directly interacting with the environment, embodying an end-to-end learning paradigm. 
	These methods leverage environmental feedback to refine decision-making policies, often employing algorithms such as Q-learning or policy gradient methods. A comprehensive review of this topic can be found in \cc{2021_SurveyAutonomousVehicle_di}.
	
	In \cite{2021_ReinforcementLearningApproach_russo}, an RL approach is presented that addresses pedestrian collision avoidance in autonomous driving systems. The focus lies on managing unexpected pedestrian crossings while maintaining trajectory tracking performance. The Deep Deterministic Policy Gradient (DDPG) algorithm is utilized to learn continuous control actions in a high-dimensional state space. The authors discuss the reward function design, agent architecture, and environment model used for training and testing the DDPG-based agent through numerical simulations. Similar data-driven concepts are presented in \cite{2014_PedMidBlock_JTTE}, \cite{2021_MotionPlanningAutonomous_rezaee}, \cite{2022_ModelingInteractionsAutonomous_trumpp} and \cite{2022_EfficientPOMDPBehavior_zhang}, demonstrating the breadth of learning-based approaches to pedestrian interaction modeling.
	Additional recent approaches have incorporated social value orientation to shape interaction policies, enabling negotiation styles from assertive to cooperative \cc{crosato2022svo}. While effective at modeling subtle dynamics, these methods remain data-intensive and challenging to verify for safety.

	However, the main drawback of these model-free concepts is that they cannot easily ensure safety or provide formal guarantees on the reliability required in high-risk environments such as AV-pedestrian interactions. The black-box nature of learned policies makes verification challenging, and their performance may degrade in scenarios not adequately represented in training data. This limitation means that pure learning-based systems are not yet mature enough to be safely deployed in real traffic situations without additional safeguards. 
	
	It is expected that model-free decision-making approaches will become easier to validate and verify for traffic admission through the combination of model-free and model-based elements, as discussed in \cite{2022_ReviewPedestrianTrajectory_korbmacher} and \cite{derajic2024learning}. Such hybrid architectures aim to leverage the adaptability of learning-based methods while retaining the interpretability and safety guarantees of model-based approaches.
	
	\subsubsection{Mechanistic Model-Based Approaches}
	In contrast, mechanistic model-based approaches commonly employ analytic criteria such as \textit{gap acceptance} and \linebreak \textit{time-to-collision} (TTC) thresholds.
	While these criteria are more straightforward to implement and provide a baseline for decision-making, a growing body of research demonstrates that human crossing behavior is richer and depends on a combination of cues. For example, \cc{tian2023deceleration} found that pedestrians rely not just on TTC, but also on the detailed parameters of vehicle deceleration when assessing whether it is safe to cross.
	
	In \cite{2022_DriverPedestrianPerceptualModels_domeyer}, a perceptual model is established based on the geometrical relations between drivers and pedestrians, emphasizing the implications for vehicle automation. It explores the coupling of driver and pedestrian behavior. However, only a simulation analysis is provided, with no strong indications for real-world usage. Game theoretical models are presented in \cite{EZZATIAMINI2024107604}, \cite{2020_AnalysisGameTheorybased_skugor} and \cite{2022_GameTheoryBasedModeling_pavelko}. In these works, the interaction between the pedestrian and vehicle is modeled as a game with two players who optimize their own objective functions through their joint actions. These models can represent the interaction between the two players, but they are not suitable for predicting their future joint actions.
	
	On the other hand, in \cite{2023_InteractionAwareDecisionMaking_chen}, a model predictive control (MPC) formulation is proposed, which can predict the future behavior of the pedestrian. In \cite{2023_StochasticModelPredictive_skugor}, the model predictive formulation is extended with a stochastic component, providing a more realistic overall behavior of the proposed algorithm. More recently, perceptually-grounded predictive models have been proposed that explicitly account for how pedestrians integrate deceleration cues and gap evolution when forming crossing decisions \cc{tian2025interacting}, and probabilistic vehicle–pedestrian interaction models have been developed to plan optimal implicit-communication behaviors under intent uncertainty \cc{amann2025optimal}.
	
	Building on these findings, perceptually plausible models have been introduced that integrate multiple sensory cues and contextual information into a continuous decision process. 
	These methods rely on a predefined model to predict future states and make decisions. 
	Furthermore, they can handle constraints and uncertainties in the system-design phase, leading to improved safety and reliability. 
	More detailed overviews of model-based methods can be found in \cite{2021_PedestrianModelsAutonomous_camara}, \cite{2021_SurveyMotionPrediction_gulzar} or \cite{2023_PedestrianBehaviorShared_predhumeau}.

	\subsection{Social Compliance and Empirical Validation}
	Interactions in traffic are multi-agent processes that combine perception, implicit signaling, and negotiation.
	Social compliance (intelligibility, predictability, comfort) is crucial for acceptance: at uncontrolled crossings AVs face a trade-off between conservative and aggressive behaviors, so decision policies must balance physical safety with legibility and social objectives \cc{tian2025tradeoff}. VR studies and human–machine interfaces (eHMI) research (text, icons, lights, kinematics) have probed trust, workload, and communication effectiveness, and work on implicit cues (e.g., head movements) shows how pedestrians infer intent \cc{2022_FactorsAffectingPedestrians_zhou,2019_HowShouldAutomated_locken,2024_InterpretingPedestriansHead_yang}. Nevertheless, many experiments constrain participant influence on AV policies, limiting mutual adaptation and leaving few studies that directly compare different decision-making algorithms under truly interactive, human-in-the-loop conditions.

	\subsection{Shortcoming of the State-of-the-Art Methods}
	
	Despite impressive progress, current state-of-the-art methods face several limitations. 
	Model-free approaches often lack interpretability and formal safety guarantees, hindering certification. 
	Model-based approaches can become computationally heavy and may not scale to dense, multi-agent scenarios, 
	and the decision-making and control algorithms from the state of the art are therefore often too complex for real-time implementation on general automotive hardware or hard to validate for motor vehicle registration. 
	Moreover, many experiments rely on scripted behaviors or do not put humans fully into the decision loop, limiting ecological validity — human in the decision making loop experiments with no predefined behavior are an inevitable extension for human factors analysis. 
	Finally, algorithms rarely consider social compliance explicitly, and there is a paucity of work on quantifying safety efficiency trade-offs. Therefore, our work addresses these challenges and provides an IAMPDM that can run in real time; furthermore, this work also provides an investigation having a human‑in‑the‑decision‑making‑loop character.	
	\section{Improvements of the Model Predictive Decision Making Algorithm} \label{sec:mpc_algo}
	\subsection{Model Description}
	The IAMPDM implements an explicit motion model for the pedestrians' motion. This is necessary to make the decision-making algorithm of the AV more suitable for city-center traffic scenarios with low driving speeds. The core idea of this IAMPDM is presented in \cc{2023_CooperativeDecisionMakingShared_varga} that we used for our implementation; and we validated the model using the data from \cc{2019_IV_Yang}.

	The proposed sigmoid-based TTC model, while simplified compared to multi-cue approaches \cc{tian2023deceleration,tian2025interacting}, captures essential gap acceptance behaviors observed in empirical studies. Research on pedestrian crossing decisions reveals that time-to-arrival estimates—closely related to TTC—serve as primary decision variables, though these estimates are systematically distorted by approach speed and individual differences \cc{2014_PedMidBlock_JTTE}. Specifically, pedestrians provide higher TTC estimates for higher approach speeds and accept smaller gaps accordingly, with age-related variations in conservativeness. Our sigmoid function's adjustable parameter $c$ accommodates these individual differences, allowing the model to represent cautious to aggressive crossing behaviors within a unified framework.
	
\begin{figure}[t!]
		\centering
		\includegraphics[width=0.8\linewidth]{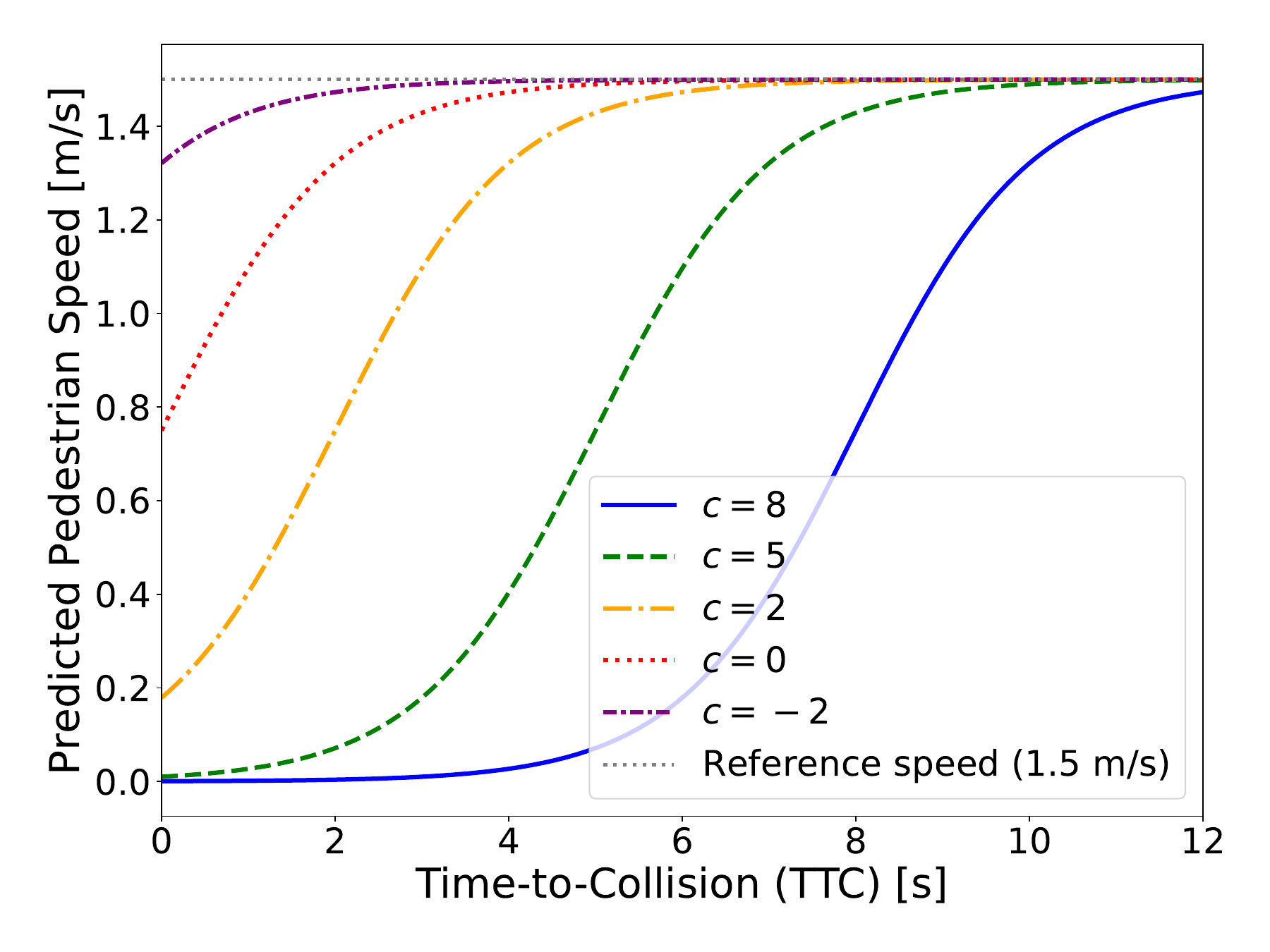}
		\caption{The output of \eqref{eq:model} with $v_\mathrm{ped}^\mathrm{ref} = 1.4 \frac{m}{a}$. Each curve represents a different pedestrian behavioral profile, with higher values of $c$ indicating more cautious decision-making and lower values reflecting assertive crossing behavior.}
		\label{fig:TTC_example}
	\end{figure}

	For the explicit motion model, it is assumed that the pedestrian's choice of speed at the next time step can be modeled by 
	\begin{equation} \label{eq:model}
		\dot{y}_\mathrm{ped}(t) = \frac{1}{1 + \exp\left(-TTC_\mathrm{MPC}(t)+c\right)} \cdot  v_\mathrm{ped}^\mathrm{ref}.
	\end{equation} 
	The impact of the parameter $c$ is illustrated in Figure \ref{fig:TTC_example}. The $TTC$ is computed as 
	\begin{equation}
		TTC_\mathrm{MPC}(t) = \frac{x_\mathrm{ped}(t) - x_\mathrm{veh}(t)}{v_\mathrm{veh}(t)} - 
		\frac{y_\mathrm{veh}(t) - y_\mathrm{ped}(t)}{v_\mathrm{ped}^\mathrm{ref}}.
	\end{equation}

	The output function of \eqref{eq:model} is a general sigmoid function and ranges between 0 and 1, which can be treated as the probability of the pedestrian crossing. The greater the $TTC$ value, the more likely it is that the pedestrian would choose to cross at a reference speed. Therefore, only a reference speed of the pedestrian $v_\mathrm{ped}^\mathrm{ref}$ needs to be identified for the model. This model allows for real-time \textit{prediction} of pedestrian motion and model-based \textit{interaction} between the pedestrian and automated vehicle, making it simultaneously suitable for practical applications. 
	
	The model's validity extends beyond its alignment with gap acceptance literature. By focusing on TTC as the core interaction metric, we follow established approaches in interaction-aware planning \cc{crosato2024social}, while our probabilistic formulation provides advantages over deterministic models. The sigmoid mapping between TTC and crossing probability naturally handles uncertainty in pedestrian intentions—a critical aspect often overlooked in rule-based systems. 
	Furthermore, validation against real-world trajectory data from \cc{2019_IV_Yang} demonstrates that this simplified model adequately captures crossing patterns in urban intersections, despite not explicitly modeling all perceptual cues that pedestrians might use.
	Importantly, the model's simplicity enables real-time computation within MPC frameworks while maintaining behavioral plausibility. This trade-off between completeness and computational efficiency reflects broader challenges in social compliance for AVs, where perfect behavioral modeling must be balanced against practical deployment constraints \cc{tian2025tradeoff}.
	
	The main benefit of the proposed model \eqref{eq:model} is that the decision layer is integrated into to control layer, thus the MPC can solve the trajectory planning and decision making tasks in one step. This can lead to a significant advantage compared to the state-of-the-art methods, where the decision-making and control layers are separated, which can lead to suboptimal solutions.
	
	Gap distance model (see e.g.~\cc{2024_DconstrHuman_tian}) inherently assumes that the pedestrian's decision-making is based on the current gap value and assessing the safety without considering how that gap might change in the future. On the other hand, \eqref{eq:model} modulates the interaction dynamics between pedestrian and vehicle, which evolves over time. It can account for variables such as speed, acceleration, and changing distances in a continuous manner and used for predicting interaction dynamics. 
	
	\begin{figure*}[t]
		\centering
		\includegraphics[width=0.85\textwidth]{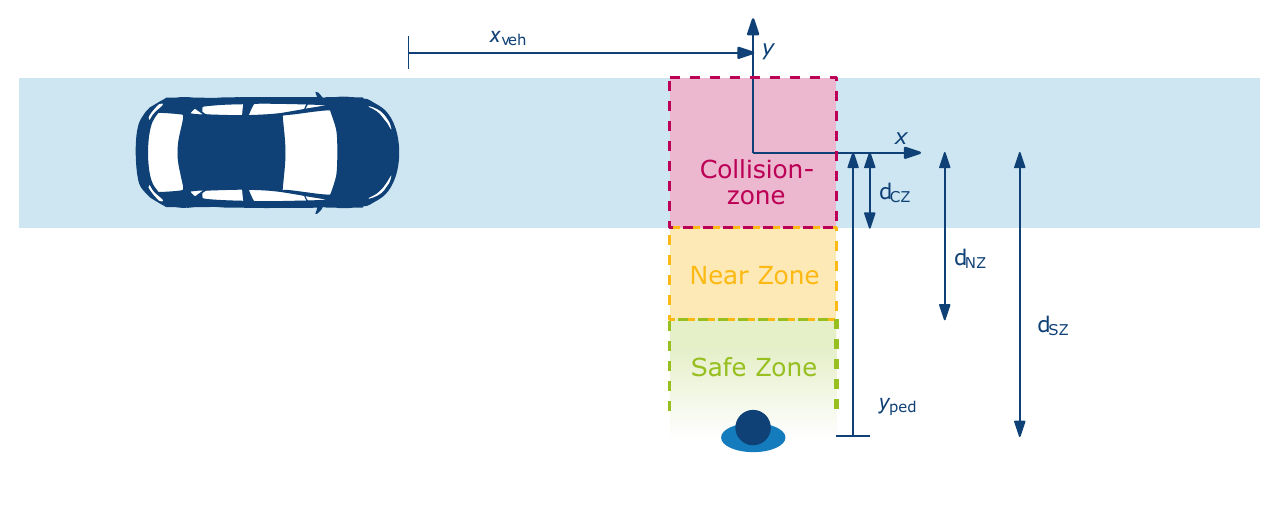}%
		\caption{Bird's eye view of the scenario with the relevant distances for decision-making}%
		\label{fig:scenario_representation}
	\end{figure*}
	
	Assuming a linear dynamics of the vehicle - which is common by the applications of autonomous vehicles - and that the pedestrian moves in the $y$ direction, the following discrete dynamic system is obtained. The state vector is defined as $\sv{x}(t) = [x_\mathrm{veh}(t), \dot x_\mathrm{veh}(t), y_\mathrm{ped}(t), \dot{y}_\mathrm{ped}(t)]^T$, where $x_\mathrm{veh}$ and $\dot x_\mathrm{veh}$ are the position and velocity of the vehicle in the driving direction, respectively. The position and velocity of the pedestrian in the $y$ direction are denoted by $y_\mathrm{ped}$ and $\dot{y}_\mathrm{ped}$, respectively. The discrete time step is $\Delta t$. The explicit dynamics of the system is given by
	\begin{align}\label{eq:explicit_dyn_model_ofMPC} \nonumber
		\underbrace{\begin{bmatrix}
				x_\mathrm{veh}(t+\Delta t) \\
				\dot x_\mathrm{veh}(t+\Delta t)\\
				y_\mathrm{ped}(t+\Delta t) \\
				\dot{y}_\mathrm{ped}(t+\Delta t)
		\end{bmatrix}}_{\sv{x}(t + \Delta t)}
		=
		\underbrace{\begin{bmatrix}
				1 & {\Delta}t & 0 &0   \\
				0 & 1 & 0 & 0  \\
				0 & 0 & 1 & \Delta t  \\
				0 & 0 & 0 & 0
		\end{bmatrix}}_{\vek{A}}
		\underbrace{\begin{bmatrix}
				x_\mathrm{veh}(t) \\
				\dot x_\mathrm{veh}(t) \\
				y_\mathrm{ped}(t) \\
				\dot{y}_\mathrm{ped}(t)
		\end{bmatrix}}_{\sv{x}(t)} \\
		+\underbrace{\begin{bmatrix}
				0.5 \cdot {\Delta}t^{2}\\
				\Delta t \\
				0 \\
				0 
		\end{bmatrix}}_{\vek{B}}  u_\mathrm{veh}(t) + \underbrace{\begin{bmatrix}
				0 \\
				0 \\
				0 \\
				\frac{v_\mathrm{ped}^\mathrm{ref}}{1 + e^{-TTC_\mathrm{MPC}(t)+c}} 
		\end{bmatrix}}_{\sv{z}(t)}
	\end{align}
	where the desired acceleration of the vehicle ${u_\mathrm{veh}(t)=a_\mathrm{des}}$ is the system input. With the model \eqref{eq:explicit_dyn_model_ofMPC}, the prediction of the future states for $N$ steps is possible using the batch formulation, see \cc{2017_PredictiveControlLinear_borrelli}. The matrices $\mathcal{A}$, $\mathcal{B}$ and $\mathcal{Z}$ are defined as:
	\begin{align*}
		\mathcal{A} =
		\begin{bmatrix}  
			\mathbf{A} \\
			\mathbf{A}^2\\
			\vdots \\
			\mathbf{A}^{N}
		\end{bmatrix} \hspace*{3mm} \mathrm{and} \hspace*{3mm} \mathcal{B} =
		\begin{bmatrix}  
			\mathbf{B} & \mathbf{0} & \ldots & \mathbf{0}\\
			\mathbf{AB} & \mathbf{B} & \ldots & \mathbf{0}\\ 
			\vdots & \ddots & \ddots & \vdots \\
			\mathbf{A}^{N-1}\mathbf{B} & \ldots & \mathbf{AB} & \mathbf{B}
		\end{bmatrix} \hspace*{2mm}
		\mathrm{and} \\
		\hspace*{3mm} \mathcal{Z} = 
		\begin{bmatrix}
			\mathbf{1} & \mathbf{0} & \ldots & \mathbf{0}\\
			\mathbf{A} & \mathbf{1} & \ldots & \mathbf{0}\\ 
			\vdots & \ddots & \ddots & \vdots \\
			\mathbf{A}^{N-1} & \ldots & \mathbf{A} & \mathbf{1}
		\end{bmatrix}.
	\end{align*}
	The prediction of the future state and input vectors are
	\begin{equation} \label{eq:pred_x}
		\vek{x}_s = \mathcal{A} \vek{x}_0 + \mathcal{B}\vek{u}_s + \mathcal{Z}\vek{z}_s,
	\end{equation}
	where for the sake of simplicity the index $s$ is used for the vector sequences: $\vek{x}_s = [\vek{x}(t), \: \vek{x}(t + \Delta t) \: \ldots \: \vek{x}(t + \Delta \cdot (N-1) t)]^T $, for the future system states, $\sv{u}_s = [u(t), \: u(t + \Delta t) \: \ldots \: u(t + \Delta \cdot (N-1) t)]^T$ for the optimizing future inputs and $\vek{z}_s = [\vek{z}(t), \: \vek{z}(t + \Delta t) \: \ldots \: \vek{z}(t + \Delta \cdot (N-1) t)]^T $ for the pedestrian future dynamics.

	\subsection{Cost Function Formulation}
	In order to formulate an MPC, the cost function with three components is defined
	\begin{equation} \label{eq:MPC_explicit_cost}
		J_\mathrm{MPC} = J_\mathrm{com} + J_\mathrm{ref} + J_\mathrm{safe},
	\end{equation}
	where three objectives are formulated: 1) comfort of the vehicle, 2) holding the reference velocity of the vehicle, and 3) safety of interaction between pedestrian and AV.	These quadratic cost functions are
	\begin{subequations} \label{eq:J_of_MPC}
		\begin{align}
			J_\mathrm{com} &= \mathrm{w}_\mathrm{com}\cdot\sv{u}^2_s \\
			J_\mathrm{ref} &= \sv{x}^\mathsf{T}_s(t)\cdot  \mathcal{Q}_\mathrm{ref} \cdot \sv{x}_s(t) \\
			J_\mathrm{safe} &= \mathrm{w}_\mathrm{safe}\cdot \frac{1} { \sv{x}^\mathsf{T}_s(t)\cdot  \mathcal{P}_\mathrm{safe} \cdot \sv{x}_s(t)},
		\end{align}
	\end{subequations} 
	where $ \mathrm{w}_\mathrm{com}$ and $\mathrm{w}_\mathrm{safe}$ design parameters. Furthermore, the matrices are
	\begin{align*}
		\mathcal{Q}_\mathrm{ref} &= \ml{diag}\underbrace{\left[\vek{Q}_\mathrm{ref},\vek{Q}_\mathrm{ref},..., \vek{Q}_\mathrm{ref}\right]}_{N \, \ml{times}} \\
		\mathcal{P}_\mathrm{safe}  & = \ml{diag}\underbrace{\left[\vek{P}_\mathrm{ref},\vek{P}_\mathrm{ref},..., \vek{P}_\mathrm{ref}\right]}_{N \, \ml{times}}
	\end{align*}
	in which for the weighting $\vek{Q}_\mathrm{ref}$ and the perturbation $\vek{P}_\mathrm{safe}$ matrices,
	\begin{align*}
		\vek{Q}_\mathrm{ref} &= \ml{diag}\left[0, \mathrm{w}_{\mathrm{ref}_\ml{veh}}, 0, \mathrm{w}_{\mathrm{ref}_\ml{ped}}\right] \, \mathrm{and} \\
		\vek{P}_\mathrm{safe} &= \ml{diag}\left[1, 0, 1, 0\right]
	\end{align*}
	hold. The decision result of the IAMPDM is obtained from the optimization  
	\begin{subequations} \label{eq:MPC_optimization}
		\begin{align}
			\sv{u}_s^* &= \mathrm{arg}\,\mathrm{min}\, J_\mathrm{MPC} \left(\sv{u}_s\right) \\
			\mathrm{s.t.}\; & \hspace*{1cm} \text{\eqref{eq:pred_x}} \\
			&d^2_\mathrm{min} \leq x_\mathrm{veh}^2(t) +  y_\mathrm{ped}^2(t) \\
			&0 \leq \dot x_\mathrm{veh}(t) \leq \dot x^\mathrm{max}_\mathrm{veh}\\
			&a_\mathrm{veh}^\mathrm{min} \leq u(t) \leq a_\mathrm{veh}^\mathrm{max},
		\end{align}
	\end{subequations}
	which computes the target acceleration of the AV. The optimization \eqref{eq:MPC_optimization} is solved by a Python API of CasADi  \cc{Andersson2019}, where the nonlinear programming solver with an interior point optimizer is applied, which can handle both the state (\ref{eq:MPC_optimization}c), (\ref{eq:MPC_optimization}d) and input constraints (\ref{eq:MPC_optimization}e). With these hard constraints, the safety of the IAMPDM can be formally ensured by strictly enforcing operational boundaries.
	
	\subsection{Integrating Intention Modeling}
	The \textit{intention} of the pedestrian to cross the street is derived from their non-verbal communication, which includes eye contact, gestures, and body posture. These \textit{explicit communication signals} are referred to as \textit{intention}. Pedestrians use these cues similarly when interacting with human-driven vehicles: They give way with hand gestures or signal their right of way by looking at the human driver. These explicit communication signals can be determined by machine-learning-based/data-driven detection systems (see e.g.~\cite{2022_CrossingNotContextBased_yang}) and should be taken into account to extend the MPC for the decision-making. Therefore, the crossing intention $I_\mathrm{ped}(t)$ of the pedestrian is introduced as a function of time and pedestrian behavior, which can vary between $0$ and $1$, which can be used for predicting the actions of the pedestrian. Note that the aim of this paper is not the development of such intention-detection algorithms. It is assumed for this work that they are given since intention-detection algorithms can be found in the literature \cc{2021_PedestrianIntentionPrediction_razali}, \cc{2022_MultiModalHybridArchitecture_rasouli}, \cc{2023_LocalGlobalContextual_azarmi}, \cc{2024_PedestrianCrossingIntention_zhou}.
	To integrate the pedestrian's crossing intention into the MPC formulation, the following two extensions are implemented: 
	\begin{itemize}
		\item[a)] If the pedestrian is in the safe or near-zone (see Figure~\ref{fig:scenario_representation}), then the parameter $\ml{w}_\ml{safe}$ of the MPC is modified based on the crossing intention of the pedestrian 
		\item[b)] If the velocity of the pedestrian in the safe or near zone is zero, a discount function for the intention is introduced. 
	\end{itemize}
	The core idea is to use the modified parameters $\ml{w}^*_\ml{safe}$ and $d^*_\mathrm{min}$ in $J_\mathrm{safe}$ by introducing
	\begin{equation} \label{eq:gain_update_based_on_intentio_w}
		\ml{w}^*_\ml{safe} =
		\begin{cases}
			\ml{w}_\ml{safe} \cdot I_\mathrm{ped}(t) & \; \text{if pedestrian is not in CZ} \\
			\ml{w}_\ml{safe} & \; \ml{else}
		\end{cases}
	\end{equation}
	and 
	\begin{equation} \label{eq:gain_update_based_on_intentio_d}
		d^*_\mathrm{min} = 
		\begin{cases}
			d_\mathrm{min} \cdot I_\mathrm{ped}(t) & \; \text{if pedestrian is not in CZ} \\
			d_\mathrm{min} & \; \ml{else}.
		\end{cases}
	\end{equation}	
	Thus, an IAMPDM is derived, which includes explicit communication elements. This extension introduces a new aspect that hasn't been applied in state-of-the-art research yet.  
	
	However, these elements can sometimes be contradictory, leading to a traffic standstill. Pedestrians often move quickly in the safe zone and slow down in the near zone. They may sometimes wait in the near zone even if they have the right of way or can cross. In such cases, an automated vehicle would stop and wait for the pedestrian\footnote{Considering the critical significance of safety, automated vehicles tend to behave conservatively, leading to more frequent stops.}. This results in neither the pedestrian nor the vehicle moving, causing a deadlock.
	As a solution, a discount function of the intention is proposed motivated by game theoretical applications, see e.g.~\cite{2021_DynamicBargainingTimeConsistency_marin-solano}:
	\begin{equation} \label{eq:discount_intention}
		I^*_\mathrm{ped}(t) = I_\mathrm{ped}(t_0) \cdot 0.9^{K_\mathrm{d}\cdot t},
	\end{equation}
	where $K_\mathrm{d}$ represents a design parameter, and $t_0$ denotes the onset of the interaction between the AV and the pedestrian. Instead of the initial crossing intention of the pedestrian $I_\mathrm{ped}(t_0)$, the discounted value is taken into account.
	
	If either the pedestrian or the vehicle has passed the intersection, no interaction occurs, allowing the vehicle to proceed at its reference speed, which is controlled by a simple velocity-tracking controller.
	
	\begin{algorithm}[t]
		\caption{The IAMPDM Algorithm with MPC and Intention Integration}
		\textbf{Input:} $ped$, $veh$, $I_\mathrm{ped}(t_0)$\\
		\textbf{Output:} $veh\_acc$\\
		
		\If{$is\_ped\_passed$ or $is\_veh\_passed$}{
			\Return{velocity\_control($ped, veh$)} 
		}
		\Else{
			\If{$is\_discounting\_intention$}{
				use (10)
			}
			\Else{
				use $I_\mathrm{ped}(t_0)$
			}
			
			\eqref{eq:gain_update_based_on_intentio_w} and \eqref{eq:gain_update_based_on_intentio_d}\\
			Update vehicle state \eqref{eq:pred_x} \\
			$veh\_acc \gets$ Optimze \eqref{eq:MPC_optimization}\\
			\Return{$veh\_acc$} 
		}

	\end{algorithm}
	Note that the main benefit of the IAMPDM is that its core structure remains the same across scenarios with few or many pedestrians. The only difference is the number of pedestrian models that need to be included in the MPC formulation. The complete IAMPDM algorithm is summarized in Algorithm~1. 
	
	\section{Experiment with the Interaction-Aware Model Predictive Decision-Making} \label{sec:experiment}
	
	In this section, the experiment and its results are presented. The goal of the experiment is to validate the applicability of the IAMPDM and compare it with 
	\begin{itemize}
		\item a non-interactive algorithm (NIA) being cautious in interactions with pedestrians and
		\item a rule-based decision-making (RBDM) introduced in\linebreak \cite{2023_IntentionAwareDecisionMakingMixed_varga}. 
	\end{itemize}
	The independent variable of the study design was the selection of the negotiation algorithm (MPDM, RBDM, NIA). For the experiment, N = 24 participants, (12 female and 12 male) were recruited from the WIVW Panel. Their mean age was 47.48 years (SD = 16.87). All participants were over 18 years old and had a valid driver’s license. A compensation of 25 Euros was paid for taking part in the experiment. The \textit{Ethical Commission of WIVW} approved the study based on \textit{Code of Ethics at WIVW – Summary for project proposals and articles/papers}.
	
	\subsection{Description of the Compared Decision-Making Algorithms}
	\subsubsection{Non-Interactive Algorithm (NIA)}
	The NIA is a conservative, non-interactive baseline that ignores explicit pedestrian crossing intentions; it applies a simple rule to stop the vehicle whenever a pedestrian is detected in the near or crossing zones (see Figure~\ref{fig:scenario_representation}) and the time-to-collision (TTC) falls below a predefined threshold. which can lead to unnecessary stops and reduced traffic-flow efficiency compared to intention-aware methods such as IAMPDM and RBDM.
	
	\subsubsection{Rule-Based Decision-Making (RBDM)}
	The RBDM is a decision-making algorithm that considers the pedestrian's intention to cross the street. The RBDM is based on a set of rules that determine the vehicle's behavior based on the pedestrian's position, speed, and intention to cross the street. The RBDM is designed to stop the vehicle if the pedestrian is in the safe zone and the TTC is less than a predefined	 threshold, but it also considers the pedestrian's intention to cross the street. If the pedestrian is in the near zone and has a high intention to cross, the RBDM will slow down the vehicle to allow the pedestrian to cross safely. The RBDM is implemented as a comparison algorithm for the IAMPDM. For more details, see \cite{2023_IntentionAwareDecisionMakingMixed_varga}.

	\subsection{Experimental Simulator Platform}
\begin{figure}[!t]
		\centering
		\includegraphics[width=0.85\linewidth]{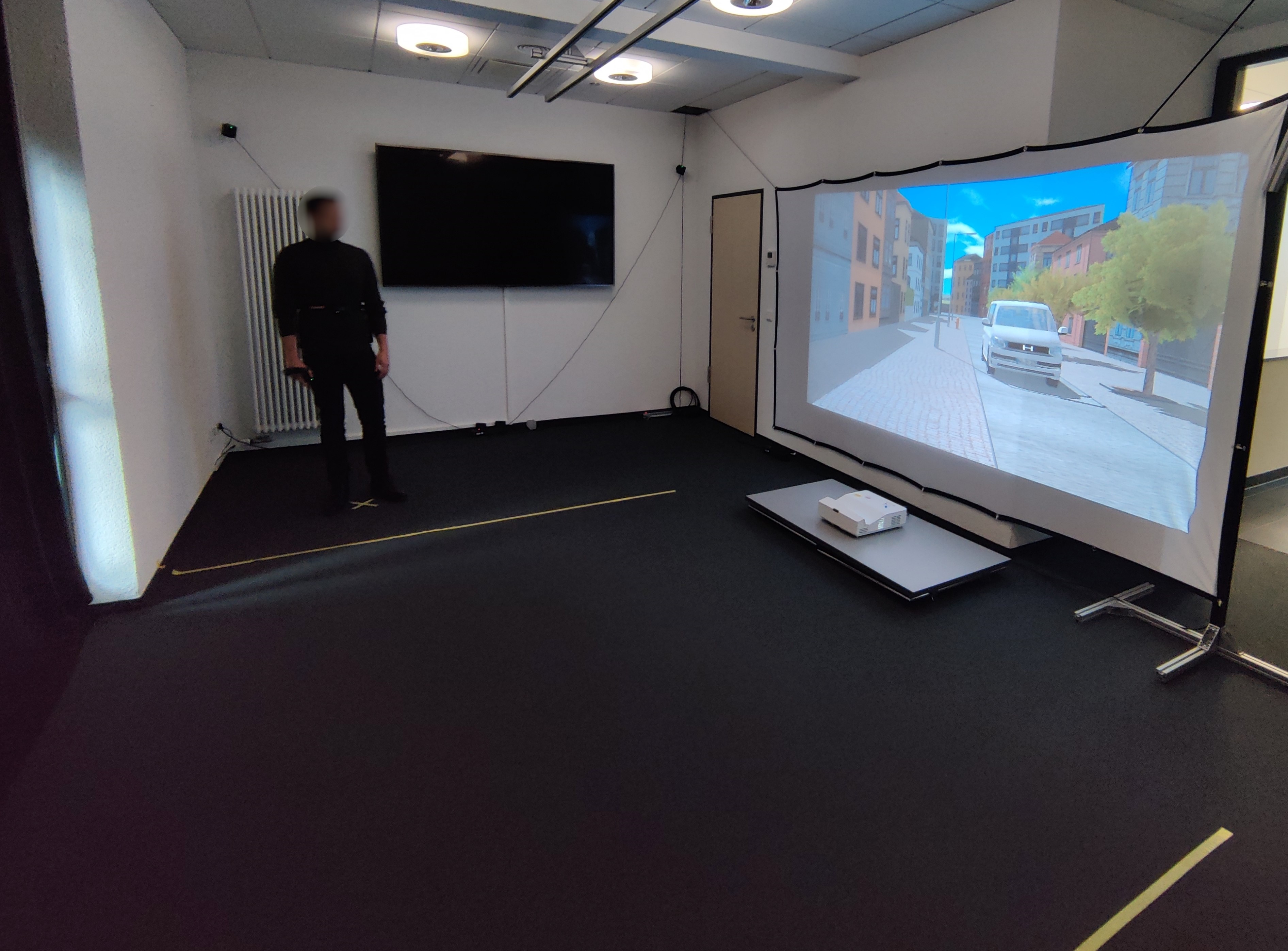}
		\caption{The simulator setup with the marked street, where the pedestrian crosess. Note that the light was turned off during the experiments, and the room was completely dark. The only light source was the projector. The test subject wore the Vive Tracker on the waist to track their position and velocity. The joystick in the test subject's hand was used to indicate their crossing intention.}
		\label{fig:proposed_architecture}
	\end{figure}

	The simulator was set up in an $8 \times 4$ meter room to simulate an urban street crossing situation for pedestrians. A screen was placed on one wall of the room, projecting the perspective of a pedestrian. This dynamic view corresponded to that of a pedestrian looking perpendicularly along the street in the crossing direction, see Figure~\ref{fig:proposed_architecture}. To precisely capture the pedestrian's position, Vive Trackers were used, with the projected view changing according to the position of the test subjects. 
	The main advantage of this configuration lies in avoiding the motion sickness of a virtual reality system, enabling longer usage and accurately representing natural crossing behavior. 
	In contrast, using a VR headset could cause simulator sickness, and the lack of visual representations of body parts in the virtual environment might lead to unnatural movement behavior. Therefore, we decided against a VR experiment based on these considerations and the advice of the ethical commission of WIVW.
	
	The experimental setup includes 
	\begin{itemize}
		\item[1] A simulation computer running SILAB$^\circledR$\footnote{SILAB is a software product of WIVW - W\"urzburg Institute for Traffic Sciences GmbH (\url{www.wivw.de})} software framework that enables real-time traffic simulation with \linebreak customizable vehicle behaviors and environmental conditions. 
		The software provides accurate physics-based vehicle dynamics and allows integration of different \linebreak decision-making algorithms for autonomous vehicles.
		
		\item[2] A short-distance projector mounted on the ceiling to display the scenario with the automated vehicle. The projector provides a large field of view (approximately 120 degrees horizontal) at a resolution of 1920x1080 pixels, enabling realistic perception of vehicle approach speeds and distances. The projection is calibrated to match real-world scale ratios.
		
		\item[3] A body sensor system consisting of HTC Vive Trackers to track the position and velocity of the test subjects in real-time with sub-centimeter accuracy. The tracking system operates at 90 Hz and provides position data of the test subjects that is synchronized with the simulation environment.
		
		\item[4] An input joystick for detecting the crossing intention of the test subjects. The joystick provides an input signal ranging from 0 to 1 that represents the strength of the crossing intention. This direct input method avoids the complexity and potential errors of computer vision-based intention recognition systems while still enabling natural interaction dynamics.
	\end{itemize}
	Using a joystick to signal crossing intention avoids computer\linebreak‑vision errors while preserving natural interaction. Since this work does not attempt to validate such detection algorithms, the use of the joystick is a reasonable solution. Therefore, the experiment is even more rigorous in its objective of comparing different decision-making algorithms in a closed-loop manner.

	\subsection{The Setups of the Decision-Making Algorithms}
	The critical design parameters of the IAMPDM are shown in Table \ref{table:parameters}. These parameters facilitate the implementation of either a more aggressive or a more conservative/cautious behavior of the AV.
	\begin{table}[t!]
		\centering
		\begin{tabular}{|c|c|}
			\hline
			Parameter & Impact of the parameter on \\
			\hline
			\hline
			$\mathrm{w}_\mathrm{safe}$ & Stopping distance\\	
			$\mathrm{w}_\mathrm{com}$ &  Deceleration rate\\
			$\mathrm{w}_{\mathrm{ref}_\ml{ped}}$&  Deceleration rate\\
			$\mathrm{w}_{\mathrm{ref}_\ml{veh}}$&  Velocity profiles\\
			$K_\ml{d}$  & Waiting time  \\
			\hline
		\end{tabular}
		\caption{The most critical parameters and their values of the IAMPDM algorithm}
		\label{table:parameters}
	\end{table}
	Instead of a manual tuning of these parameters, it is beneficial to apply an automated, optimization-based tuning framework. However, the usage of such automated tuning frameworks from the literature was not suitable for setting up the IAMPDM since they do not take human subjective preferences into account. The inclusion of human preferences into IAMPDM tuning is one challenging aspect since it helps to enhance public acceptance of such technical systems. 
	
	An automated tuning framework for pedestrian-AV interaction is presented in~\cite{2023_IntentionAwareDecisionMakingMixed_varga}, which utilizes the optimization
	\begin{subequations} \label{eq:Parameter_optimization}
		\begin{align}
			\sv{\theta}^* &= \mathrm{arg}\,\mathrm{min}\, J_\mathrm{glob} \\
			\mathrm{s.t.}\; & \eqref{eq:MPC_optimization}
		\end{align}
	\end{subequations}
	to find the optimal parameter vector $\sv{\theta}$ of an intention-aware decision-making. The global cost function has the form
	\begin{align} \label{eq:f_PSO} \nonumber
		J_\mathrm{glob} = \int_{t_0}^{T_\mathrm{end}}
		k_1 \cdot t + k_2 \cdot \left|a^2_\mathrm{max,veh}(t)\right| \\
		- k_3 \cdot \left|d_\mathrm{min}\right| + k_4 \cdot \frac{1}{TTC_\mathrm{MPC}(t)} \,\mathrm{d}t, 	
	\end{align}
	from which the optimal parameters of the intention-aware decision-making are obtained. In previous works, the parameters $k_i, i = \{1,2,3,4\}$ were determined based on optimization goals, such as acceleration profiles and stopping distances. These selections, however, do not account for human preferences. Therefore, for this work, we suggest a semi-automated expert design process to systematically tune the parameters $k_i$. The proposed semi-automated expert design of the IAMPDM has the following steps:
	\begin{algorithm}[t]
		\caption{The algorithm of the iterative design of the intention-aware decision-making including human preferences}\label{alg:IAA}
		\KwIn{$k_i,  i = \{1,2,3,4\}$}
		\KwOut{$\sv{\theta}^*$}
		
		\While{Overall results not sufficient}{
			Setting $k_i,  i = \{1,2,3,4\}$\\			
			Run \eqref{eq:Parameter_optimization} with $J_\ml{glob}$\\
			Testing the decision-making\\	
		}
	\end{algorithm}
	
	Through this iterative design, we can obtain the optimal parameters for IAMPDM integrating human preferences. By including these preferences, we achieve more human-centered optimization results. This human-centered optimization is beneficial because it requires tuning only a few parameters, no matter the algorithm's complexity, since the parameter vector of IAMPDM consists of 
	$$
	\sv{\theta}=\left[\mathrm{w}_\mathrm{safe}, \mathrm{w}_\mathrm{com}, \mathrm{w}_{\mathrm{ref}_\ml{ped}}, \mathrm{w}_{\mathrm{ref}_\ml{veh}}, d_\mathrm{min}, K_\ml{d}, x^\mathrm{max}_\mathrm{veh}, a_\mathrm{veh}^\mathrm{min}, a_\mathrm{veh}^\mathrm{max}\right]
	$$
	that can be difficult to tune manually. Moreover, $k_i, i = {1,2,3,4}$ always preserves their physical meaning, leading to time-efficient human-centered optimization. 
	
	The RBDM is also tuned with Algorithm \ref{alg:IAA}; the only difference is that the RBDM has fewer parameters. To mitigate the risk of injury in urban areas, the non-interactive algorithm assumes that pedestrians near the street edge may cross unpredictably at any time. As a result, the vehicle halts and waits for a duration of $t_\ml{NIA}$, disregarding pedestrian crossing intentions since it assumes the absence of high-level communication detection. If the pedestrian does not cross, the vehicle starts moving again cautiously after $t_\ml{NIA}$, in accordance with the IAMPDM discount function, which leads to similar wait times.

	\subsection{Experiment Design}
	The three different decision\verb|-|making algorithms were \linebreak marked with different colors during the experiment, meaning that they were not revealed to the test subjects. As a result, distinguishing was simplified without using numbers, resulting in fewer matching errors by the test subjects.
	
	Since existing research does not address the \textit{human-in-the-decision-loop} nature of pedestrian-AV interactions in unsignalized crossings, a key challenge was to instruct test subjects in a way that 1) facilitates interaction but 2) does not result in a fixed sequence of human actions that entirely predetermines their behavior. This balance is challenging, as too much freedom leads to high variability, whereas fully predefined behavior compromises the evaluation of the IAMPDM. Forcing the interaction may result in unrealistic scenarios and misleading results.
	
	This challenge was addressed by instructing the test subjects to mimic altering their decisions: They started to cross the street, but they waited on the roadside. Therefore, we defined four \textit{scenarios}, which describe approximately how the test subjects should behave, but no fixed sequence of human actions is defined, no crossing time and distance specifications for specific actions are given in advance. The detailed instructions are given in the Appendix. These \textit{scenarios} are
	\begin{itemize}
		\item[1] \textit{Crossing} before the vehicle
		\item[2] \textit{Remaining} and letting the vehicle to cross first  
		\item[3] \textit{Delayed Crossing} before the vehicle
		\item[4] \textit{Delayed Remaining} and letting the vehicle to cross first.
	\end{itemize}
	
	The experimental procedure was organized as follows: \linebreak Test subjects started with scenarios 1 and 2, performed in a randomized order. Each scenario involved three crossings using each of the three decision-making algorithms, leading to 9 crossings per scenario. After completing both scenarios, they filled out an \textit{intermediate questionnaire} to assess the algorithms. This questionnaire was crucial for shaping the test subjects' perspectives for the final evaluation.
	
	Afterward, the test subjects repeated the procedure with scenarios 3~and~4 in a randomized sequence. They tested all the decision-making algorithms and filled out the \textit{intermediate questionnaire} again. Finally, they had to answer the \textit{final questions}, which are used for subjective assessment of the algorithms. Since scenarios 3 and 4 are the most crucial and involve contradictory interactions with the AV, only these were used for objective assessment.

	\subsection{Objective Goals and Evaluation Criteria}
	In the following, our null and working hypotheses are presented, and the measures of the experiment are discussed. 
	
	The investigation included two hypotheses focusing on objective and subjective measures. For the objective assessment, we have the following null and working hypotheses:
	\begin{itemize}
		\item[H1$_0$] The use of intention‑aware decision‑making does not significantly alter the crossing times and the criticality of intersection scenarios.
		\item[H1$_w$] The use of intention-aware decision-making in autonomous vehicles significantly shortens crossing times and elevates the criticality of intersection scenarios.
	\end{itemize}
	The subjective assessment is based on the following null and alternative hypotheses:
	\begin{itemize}
		\item[H2$_0$] The usage of intention-aware decision-making does not lead to a higher user preference in the intersection scenarios.
		\item[H2$_w$] The usage of intention-aware decision-making leads to a higher user preference in the intersection scenarios.
	\end{itemize}
	To measure our hypotheses, we used three evaluation metrics defined in~\cite{2023_CriticalityMetricsAutomated_westhofen}: 
	\begin{itemize}
		\item The average Time-to-Collision ($TTC_\ml{avg}$). $TTC(t)$ is computed such as
		\begin{equation}
			TTC(t) = \frac{y_\ml{ped}(t) + x_\ml{veh}(t)}{\ml{max}(v_\ml{veh}(t),\kappa)},
		\end{equation}
		where $\kappa = 0.05$ in order to ensure the numerical stability, in cases of the pedestrian or the vehicle are at a standstill. From that the average $TTC_\ml{avg}$ is calculated by
		\begin{equation}
			TTC_\ml{avg} = \frac{1}{N} \int_{t_0}^{T_\ml{end}} TTC(t) \, \text{d}t. 
		\end{equation}
		\item Average Deceleration to Safety Time (DST) 
		\begin{equation}
			DST(t) = \frac{1}{2}\frac{v^2_\ml{ped}(t) + v^2_\ml{veh}(t)}{x_\ml{veh}(t) + y_\ml{ped}(t) + v_\ml{veh}(t)\cdot t_\ml{safe}},
		\end{equation}
		where the safety time is $t_\ml{safe}=1$. The average $DST_\ml{avg}$ is calculated by
		\begin{equation}
			DST_\ml{avg} = \frac{1}{N} \int_{t_0}^{T_\ml{end}} DST(t) \, \text{d}t. 
		\end{equation}
		\item The completion time of the scenario ($T_\ml{end}$), defined as the time after the vehicle or pedestrian has left the collision zone.
	\end{itemize}
	Besides these objective metrics, the test subjects had to assess the decision-making algorithm in the \textit{final questions} to obtain their subject impressions. These are
	\subsubsection*{Question 1:} \textit{Which decision-making did you find overall to be the most appropriate concerning the experienced situations?}
	
	\subsubsection*{Question 2:} \textit{Please now evaluate the decision-making algorithms based on your experiences in all the situations: Can you imagine a red/green/blue vehicle with this decision-making algorithm operating in real traffic?}\\
	\textit{What do you think of red/green/blue decision-making in general?}
	
	The possible answers are given in Table \ref{tab:answer_table}.
	\begin{table}[!h]
		\centering
		\begin{tabular}{|*{6}{@{\hspace{4pt}}c@{\hspace{4pt}}|}} 
			\hline
			\begin{tabular}{@{}c@{}}not \\ at all\end{tabular} & \begin{tabular}{@{}c@{}}very \\ little\end{tabular} & little & medium & strong & \begin{tabular}{@{}c@{}}very \\ strong\end{tabular}  \\
			\hline
			0&1,2,3 &4,5,6 &7,8,9 &10,11,12 &13,14,15 \\
			\hline
		\end{tabular}
		\caption{Answer Table for the test-subjects}
		\label{tab:answer_table}
	\end{table}
	
	Note hypothesis H1 is evaluated by the objective metrics $TTC$, $DST$ and $T_\ml{end}$. For the assessment of hypothesis H2 the final questions are used.

	\section{Results} \label{sec:Results_discussion}
	This section presents the results of our experiment and provides an in-depth discussion. Please note that the \textit{Delayed Crossing} and \textit{Delayed Remaining} scenarios cause different overall behaviors in the pedestrian and the automated vehicle, making it necessary to assess them separately. For both objective and subjective results, the outliers are removed using the inter-quartile range method, see~e.g.~\cite{2023_OutliersDetectionElimination_dash}.
	\subsection{Objective Results}
\begin{figure*}[t!]
		\centering
		\begin{tabular}{ccc}
			\begin{minipage}{0.3\textwidth}
				\centering
				\includegraphics[width=\textwidth]{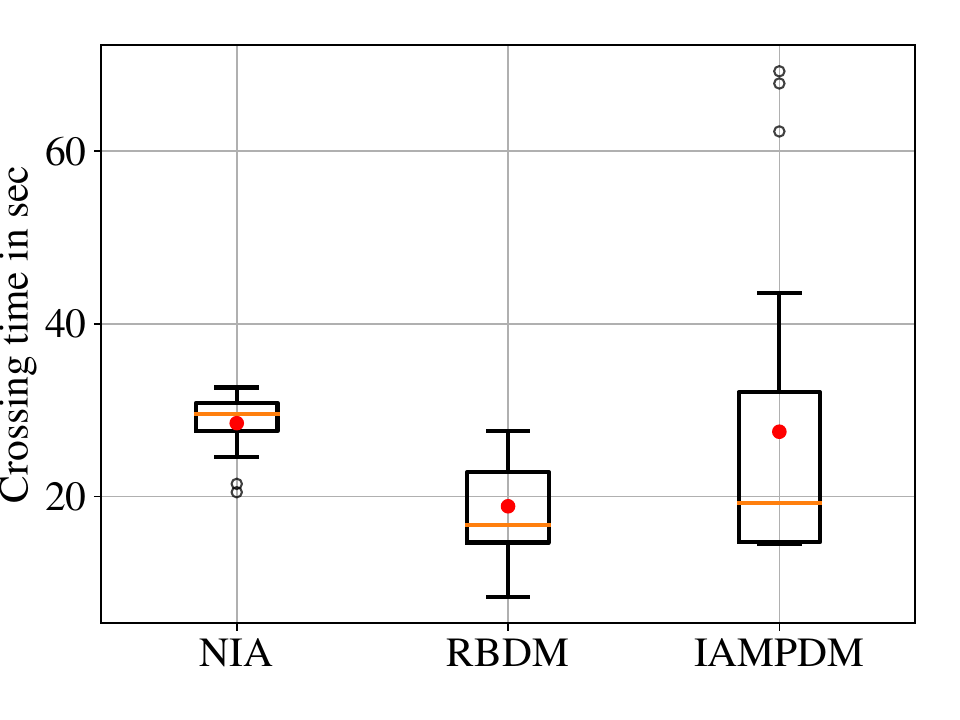} 
				\caption*{(a) Average crossing times}
				\label{fig:sub1} 
			\end{minipage}
			&
			\begin{minipage}{0.3\textwidth}
				\centering
				\includegraphics[width=\textwidth]{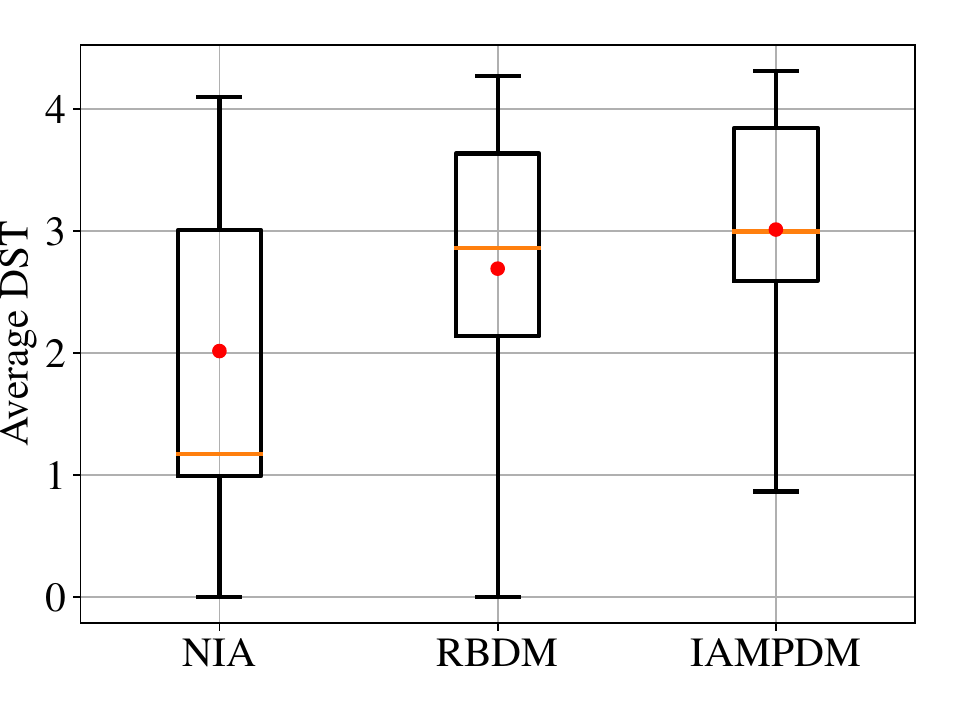} 
				\caption*{(b) Average DSTs}
				\label{fig:sub2} 
			\end{minipage}
			&
			\begin{minipage}{0.3\textwidth}
				\centering
				\includegraphics[width=\textwidth]{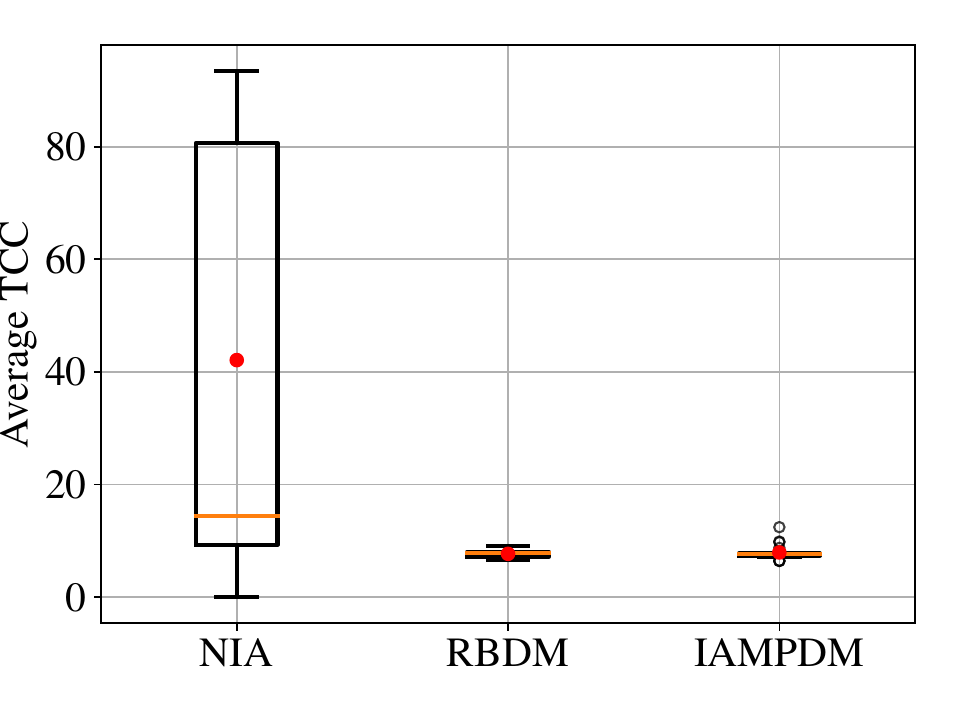} 
				\caption*{(c) Average TTCs for \textit{Delayed Crossing}}
				\label{fig:sub3} 
			\end{minipage}
		\end{tabular}
		
		\caption{Results in case of \textit{Delayed Crossing}}
		\label{fig:box_plots_delayed_crossing}
	\end{figure*}

\begin{figure*}[t!]
		\centering
		\begin{tabular}{ccc}
			\begin{minipage}{0.3\textwidth}
				\centering
				\includegraphics[width=\textwidth]{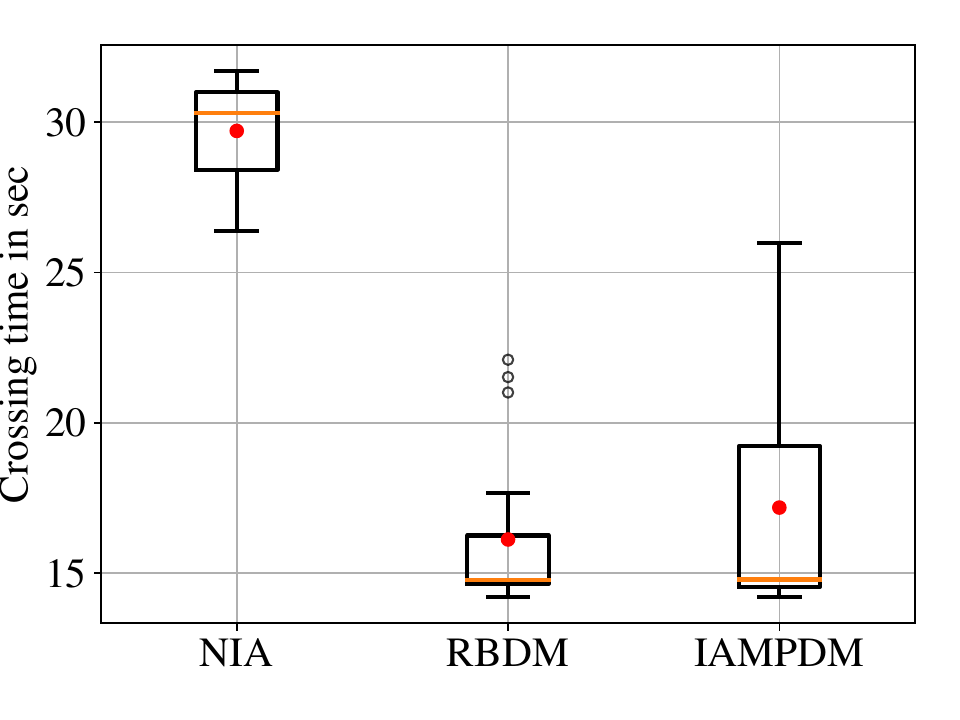} 
				\caption*{(a) Average crossing times}
				
			\end{minipage}
			&
			\begin{minipage}{0.3\textwidth}
				\centering
				\includegraphics[width=\textwidth]{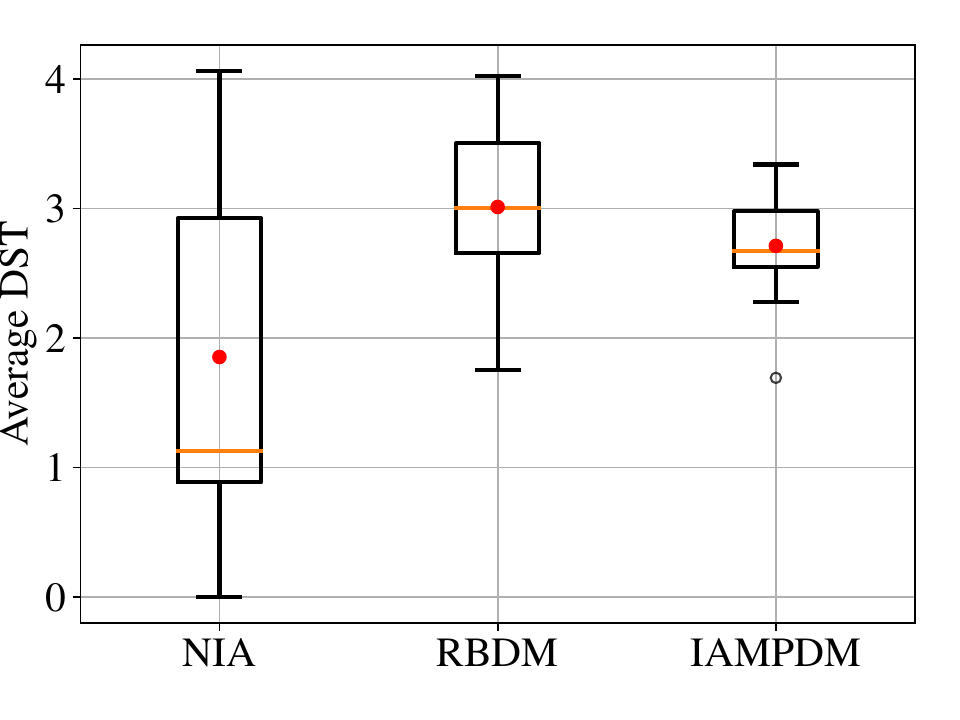} 
				\caption*{(b) Average DSTs}
				
			\end{minipage}
			&
			\begin{minipage}{0.3\textwidth}
				\centering
				\includegraphics[width=\textwidth]{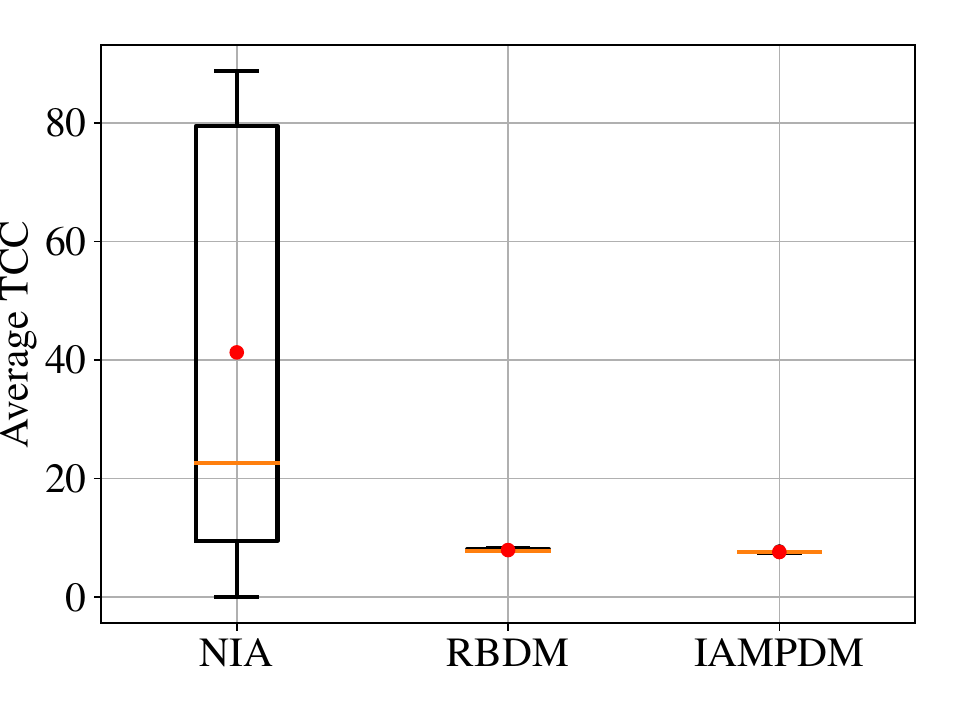} 
				\caption*{(c) Average TTCs for \textit{Delayed Remaining}}
				
			\end{minipage}
		\end{tabular}
		
		\caption{Results in case of \textit{Delayed Remaining}}
		\label{fig:box_plots_delayed_no_crossing}
	\end{figure*}

	The resulting average values of the measures $TTC_\ml{avg}$, \linebreak $DST_\ml{avg}$ and $T_\ml{end}$ with their standard deviations are presented in Table~\ref{tab:res_crossing} for the \textit{Delayed Crossing} case and in Table~\ref{tab:res_remaining} for the \textit{Delayed Remaining} case. Furthermore, Figure~\ref{fig:box_plots_delayed_crossing} and Figure~\ref{fig:box_plots_delayed_no_crossing} show the box plots of the objective results.	
	
	It can be seen that the NIA has the largest $T_\ml{end}$ for both scenarios. On the other hand, the IAMPDM and RBDM have comparable shorter crossing times. This means that negotiations between a pedestrian and an AV took shorter, indicating that using intention-aware decision-making algorithms can lead to smoother traffic.
	
	On the other hand, the criticality metrics (DST and TTC) show that the scenarios became more critical: The TTC was smaller for both IAMPDM and RBDM compared to NIA. Furthermore, the DST values were smaller if the NIA was used\footnote{Note that the smaller the DST, the less critical the scenario. On the other hand, larger TTC values mean safer (less critical) scenarios.}. 
	
	To assess whether these differences between the decision-making algorithms are statistically significant, statistical tests are performed. First, the Kruskal-Wallis test is utilized for evaluating H1, because all three samples are compared, see e.g.~\cite{2008_IntroductoryStatistics_dalgaard}.
	The degrees of freedom of this test are $df=2$, and the significance level is chosen to $\alpha=0.01.$ Its null hypothesis is that there is no difference between the three decision-making algorithms. This hypothesis is declined if $\mathcal{H} \geq \mathcal{X}_{df,\alpha}^2$ holds, where $\mathcal{X}^2_{df=2,\alpha=0.01} = 9.21$.
	
	\input{res_table_v2crossing.tex}
	\input{res_table_v2not_crossing.tex}
	In case of \textit{Delayed Crossing}, the following $\mathcal{H}$ values are obtained
	\begin{align*}
		\mathcal{H}_{Crs_{T_\ml{end}}} &= 16.64 \\
		\mathcal{H}_{Crs_{DST}} &=  6.82\\
		\mathcal{H}_{Crs_{TTC}} &= 22.42.\\
	\end{align*}
	Since $\mathcal{H}_i \geq \mathcal{X}_{df,\alpha}^2$ hold for $i = \{Crs_{T_\ml{end}, Crs_{TTC}}\}$, we can reject the null hypothesis H1$_0$ for the \textit{Delayed Crossing} scenarios that the average $T_\ml{end}$ and $TTC$ are same for all three decision-making. This indicates that the choice of the decision-making algorithms leads to statistically significant differences in $T_\ml{end}$ and $TTC$. On the other hand, $\mathcal{H}_i \geq \mathcal{X}_{df,\alpha}^2$ does not hold for $i = Crs_{DST}$ meaning the we cannot reject the null hypothesis of H1 in case of $DST$. In case of \textit{Delayed Remaining}, the following $\mathcal{H}$ values are obtained
	\begin{align*}
		\mathcal{H}_{Rem_{T_\ml{end}}} &= 32.44\\
		\mathcal{H}_{Rem_{DST}} &=9.91 \\
		\mathcal{H}_{Rem_{TTC}} &= 30.09\\
	\end{align*}
	Since $\mathcal{H}_i \geq \mathcal{X}^2_{df,\alpha}$ is satisfied for  {$i = \{Rem_{T_\ml{end}}, Rem_{DST}, Rem_{TTC}\}$}, we can conclude that the choice of decision-making algorithms has statistically significant effects on $T_\ml{end}$, $DST$ and $TTC$. In addition, Mann–Whitney–Wilcoxon tests are carried out to compare IAMPDM and RBDM. As it can be seen from Table~\ref{tab:p_mpc_rb}, aside from $TTC$ in \textit{Delayed Remaining} case, there is no statistical difference between IAMPDM and RBDM. 

	\begin{table}[t!]
		\centering
		\begin{tabular}{|l|c|} 
			\hline
			Parameter & $p$-value \\
			\hline \hline
			$p_\mathrm{Crs_{T_\mathrm{end}}}$ & \num{0.34} \\
			$p_\mathrm{Crs_{DST}}$ & \num{0.45} \\
			$p_\mathrm{Crs_{TTC}}$ & \num{0.51} \\
			\hline \hline
			$p_\mathrm{Rem_{T_\mathrm{end}}}$ & \num{0.73} \\
			$p_\mathrm{Rem_{DST}}$ & \num{0.08} \\
			$p_\mathrm{Rem_{TTC}}$ & \num{9.47e-6} \\
			\hline
		\end{tabular}
		\captionof{table}{The resulting $p$-values of Mann–Whitney–Wilcoxon tests to compare IAMPDM and RBDM}
		\label{tab:p_mpc_rb}
	\end{table}

	\subsection{Subjective Results}
	For the testing of H2, the results of the \textit{final questions} are analyzed, for which the test subject Nr. 21 is excluded since the person could not recall which decision-making was which. First, the preferences of the test subject are given in Table \ref{tab:Q1_final_quesitions}. It can be seen that most of the test subjects have chosen intention-aware decision-making. The results of Q2 from the \textit{final questions} are given in Table~\ref{tab:final_question_table}. The raw data of Q2 is included in~\ref{app2}. It can be seen that in the subjective assessment, both RBDM and the IAMPDM possess larger mean values than NIA. The Kruskal-Wallis test is conducted to test the difference for statistical significance. The result is
	$
	\mathcal{H}_\ml{Subj} = 14.56,
	$
	which indicates that the choice between intention-aware and non-interactive decision-making algorithms leads to statistically significant differences in the preference of the human test subjects. For a comparison between the IAMPDM and RBDM, a Mann–Whitney–Wilcoxon test is carried out, which yields $p_\ml{Subj}=0.154,$ indicating that the difference between IAMPDM and RBDM is not significant.
	
	\begin{table*}[h!]
		\centering
		\begin{tabular}{cc}
			\parbox{0.45\textwidth}{
				\centering
				
				\begin{tabular}{|c|c|}
					\hline
					& Preferences \\
					\hline
					NIA & 5 \\
					RBDM & 14 \\
					IAMPDM & 5 \\
					\hline
				\end{tabular}
				\captionof{table}{Preferences of the test subjects based on Q1}
				\label{tab:Q1_final_quesitions}
			}
			&
			\parbox{0.45\textwidth}{
				\centering
				\begin{tabular}{|l|c|}
					\hline
					& \begin{tabular}{c}Final \\ Question Score\end{tabular} \\
					\hline
					$\mu_{\mathrm{NIA}}$ & 7.04 \\ 
					$\sigma_{\mathrm{NIA}}$ & 3.629 \\
					\hline
					$\mu_{\mathrm{RBDM}}$ & 11.04 \\ 
					$\sigma_{\mathrm{RBDM}}$ & 3.665 \\
					\hline
					$\mu_{\mathrm{IAMPDM}}$ & 10.00 \\ 
					$\sigma_{\mathrm{IAMPDM}}$ & 2.690 \\
					\hline
				\end{tabular}
				\captionof{table}{Mean values and standard deviations of Q2}
				\label{tab:final_question_table}
				
			}
		\end{tabular}
	\end{table*}

	\begin{figure}[!t]
		\centering
		\includegraphics[width=0.7\linewidth]{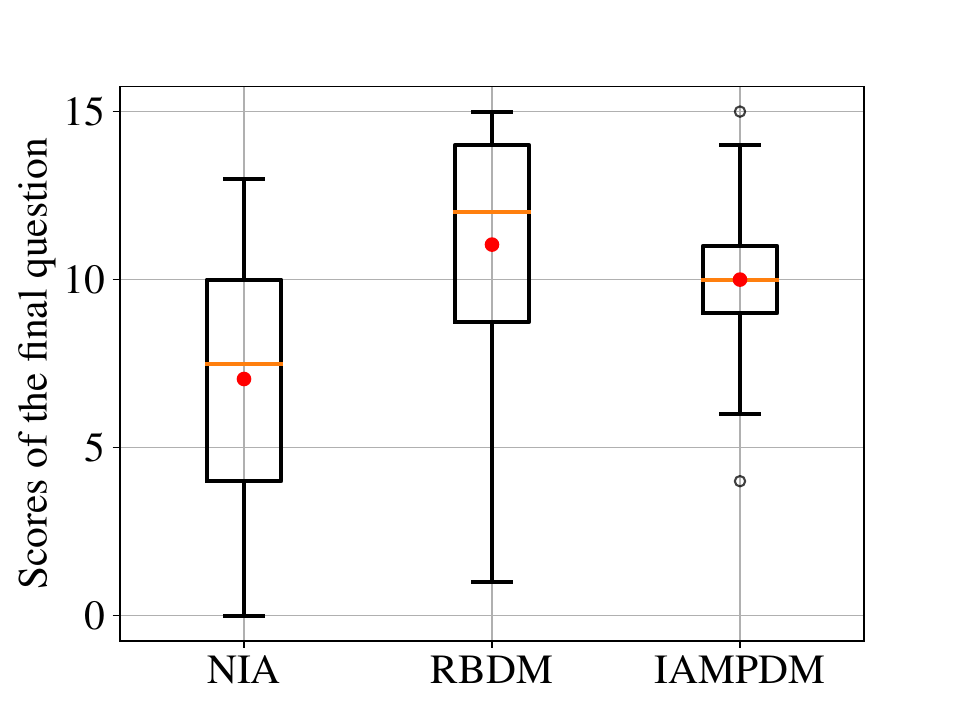}
		\caption{Results of the subjective assessment}
		\label{fig:final_question_plot}
	\end{figure}
	
	\subsection{Discussion and Limitations}
	
	After presenting the results, we discuss the implications and limitations of this study. Firstly, it is necessary to highlight that higher TTC values are associated with natural driving behavior of the AV. This preference is reflected in the test subjects choosing the RBDM and IAMPDM over the more conservative NIA algorithm. Additionally, the test subjects provided insights into the general question \textit{What do you think of red/green/blue decision-making in general?}, which enhanced our understanding of the proposed algorithms.
	
	From the study results, we can draw interesting conclusions. A closer examination of the objective results reveals that the intention-aware decision-making algorithms consistently outperform the NIA in all aspects. However, there is no significant difference between IAMPDM and RBDM, except in the average TTCs concerning \textit{Delayed Remaining} (see Table~\ref{tab:p_mpc_rb}), where IAMPDM exhibited slightly superior performance. For the other cases and measures, no significant differences between IAMPDM and RBDM were observed in this study.
	
	Additionally, we observed sporadic instances of extremely high crossing times for IAMPDM in Figure~\ref{fig:box_plots_delayed_no_crossing}(a). These outliers primarily occurred when pedestrians hesitated mid-crossing, triggering repeated velocity adjustments of the AV. This reactive behavior suggests the need for better anticipation of pedestrian motion variability during the prediction, see e.g. \cc{tian2025interacting}.
	
	Road traffic interactions involve reciprocal behaviors where agents adapt based on mutual observations.
	While \cite{2023_InteractionAwareDecisionMakingAutomated_crosato} and \cite{2020_DefiningInteractionsConceptual_markkula} use dynamical models that integrate social awareness for decision\,-making, our work aligns with their frameworks by modeling pedestrian intentions within an MPC structure.
	
	While our work utilizes TTC as a core interaction metric, it is recognized that pedestrians rely on a multitude of cues, see e. g.  \cite{tian2023deceleration}.
	Furthermore, models like the one proposed by~\cite{tian2025interacting} validate this multi-cue hypothesis. Our IAMPDM can be extended with these additional cues in future work.
	
	The results suggest that a simple algorithm, like the RBDM, can achieve outcomes comparable to more complex methods in \textit{simple} urban traffic scenarios. This implies that distinguishing between \textit{simple scenarios} and \textit{complex scenarios} is necessary for the validation of interaction-aware algorithms for autonomous vehicles. This distinction is crucial because while rule-based systems like RBDM suffice for simple cases, complex scenarios demand the capabilities of model-based approaches like IAMPDM. RBDM is a heuristic algorithm that adjusts velocity based on TTC and intention thresholds (e.g., yielding if pedestrian intention exceeds 0.5), monitoring pedestrian position and velocity. In contrast, IAMPDM employs a predictive model to optimize vehicle trajectories, considering pedestrian intentions and future states. This allows IAMPDM to anticipate pedestrian actions and plan accordingly, which is particularly beneficial in complex scenarios with multiple agents or ambiguous situations.
	
	Moreover, the subjective results offer interesting insights from the human-in-the-decision-loop experiment. The test subjects who preferred the NIA indicated that a more conservative approach could increase safety and be more favored by pedestrians. They had comments such as:
	\begin{itemize}
		\item ``\textit{Too passive and slow for my taste, but that's not wrong in normal traffic. Then misunderstandings are not so tragic.}''
		\item ``\textit{As a pedestrian, I had the feeling that it [the vehicle] looked out for me in every situation and always gave me the opportunity to cross the road. Overall, I felt safest with [NIA].}''
		\item ``\textit{Decision-making for the pedestrians pleasant.}''
	\end{itemize}
	
	Most of the test subjects preferred RBDM over the \linebreak IAMPDM, see Table~\ref{tab:Q1_final_quesitions}. However, there is no significant difference in the objective results; see Table~\ref{tab:final_question_table}. Furthermore, most of them were not able to distinguish between RBDM and IAMPDM, which are reflected in comments such as
	\begin{itemize}
		\item ``\textit{I found hard to the difference between [RBDM] and} \; \textit{[IAMPDM], but [RBDM] seemed a bit more thoughtful. Reacted better and more appropriately than [IAMPDM].}''
		\item ``\textit{[In case of IAMPDM:] To [RBDM] no difference detectable}''
		\item ``\textit{[In case of RBDM:] no big difference to [IAMPDM]}''
	\end{itemize}
	
	The reason for not detecting a significant difference between RBDM and IAMPDM is assumed to be the tuning framework. 
	Both IAMPDM and RBDM were tuned with the same semi-automated, expert-in-the-loop process, which may reduce separability between methods and introduce bias toward smoothness and comfort. The NIA baseline relies on a waiting-time parameter that influences both safety margins and delay.
	
	Some general comments were that the test subjects were not sure whether the vehicle recognized them or not. To overcome this, implementing an eHMI for a vehicle to communicate with pedestrians could significantly enhance trust. When a pedestrian sees a clear signal from the vehicle indicating it has detected them, it reduces uncertainty and potential accidents.
	
	Results indicate that increased algorithmic complexity does not improve performance or user preference in simple, constrained scenarios. Preassumbly, advanced models like \linebreak IAMPDM require targeted validation in more complex environments with larger interaction spaces to demonstrate their benefits. Future work will include additional metrics and testing in multi-agent, ambiguous settings to further explore these dynamics.
	
	The resulting practical implications for policy and deployment are that policymakers should set naturalness-aligned safety floors (e.g., min-TTC gaps and deceleration bounds), require conservative fallback when intention certainty is low, and mandate dual-objective calibration with fail-soft timeouts to prevent indecision oscillations.

	\section{Summary} \label{sec:summary}
	
	This paper proposes an interaction-aware model predictive decision-making system and its real-time implementation for \linebreak automated vehicles interacting with pedestrians in urban, low-speed scenarios. Furthermore, we designed a human-in-the-decision-loop study to examine the human-automation action-reaction cycle. 
	The human-in-the-decision-loop study,\linebreak conducted with 25 participants, demonstrated that intention-aware decision-making algorithms lead to faster resolution of human-automation negotiations and improved subjective evaluations. These findings highlight the significance of intention-aware decision-making for autonomous vehicles in urban settings.
	Our future work will feature an adaptive decision-making algorithm based on~\cite{2024_AdaptiveCooperationModelBased_varga}. Furthermore, we plan to integrate a probabilistic estimation of human crossing intentions by modeling their stochastic behavior, see \cite{2023_StochasticModelPredictive_skugor}.

	\section*{CRediT authorship contribution statement}
	\textbf{Balint Varga:} Writing – original draft, Writing – review \& editing, Conceptualization, Formal analysis, Methodology, Software, Visualization, Validation.
	\textbf{Thomas Brand:} Writing – review \& editing, Conceptualization, Data curation, Methodology, Software.
	\textbf{Marcus Schmitz:} Writing – review \& editing, Methodology, Conceptualization.
	\textbf{Ehsan Hashemi:} Writing – review \& editing, Investigation, Conceptualization.
	
	\section*{Funding sources}
	This work was supported by the Federal Ministry for Economic Affairs and Climate Action, in the New Vehicle and System Technologies research initiative with Project number \linebreak 19A21008D.
	
	\section*{Declaration of competing interest}
	The authors declare that they have no known competing financial interests or personal relationships that could have appeared to influence the work reported in this paper.
	
	\section*{Declaration of generative AI and AI-assisted technologies in the manuscript preparation process}
	During the preparation of this work, the authors used ChatGPT (OpenAI) to assist with improving language, readability, and organization of parts of the manuscript. After using this tool, the authors reviewed and edited the content as needed and take full responsibility for the content of the published article.


	\bibliographystyle{elsarticle-num}

\input{varga_IAMPDM_fin_2026.bbl}
	\newpage
	\appendix
	\section{Instruction Materials} \label{app1}
	\subsection*{Crossing}
	“Please proceed to the markings. In this situation, you want to cross the road in front of the vehicle. Behave as you would with a real vehicle and driver. Please only interact with the vehicle when you think the vehicle is aware of you. Remember to press the button when you want to signal that you want to cross the road. Keep the button pressed as long as you want to cross the road in front of the vehicle. Please remember that you should only cross the road if you feel safe, otherwise do not cross the road.
	
	\subsection*{Remain}
	“Please proceed to the markings. In this situation, you only want to cross the road after the vehicle. Behave as you would with a real vehicle and driver. Remember to press the button when you want to signal that you want to cross the road. Keep the button pressed as long as you want to cross the road in front of the vehicle.”
	
	\subsection*{Delayed crossing}
	“Please proceed to the markings. In this situation, you actually want to cross the road after the vehicle. However, you change your mind and spontaneously decide to cross the road before the vehicle. Behave as you would with a real vehicle and driver. Please only interact with the vehicle when you think the vehicle is aware of you. Remember to press the button when you want to signal that you want to cross the road. Keep the button pressed as long as you want to cross the road in front of the vehicle. Please remember that you should only cross the road if you feel safe, otherwise do not cross the road.”
	
	\subsection*{Delayed stay}
	“Please move to the markings. In this situation, you actually want to cross the road in front of the vehicle. However, you change your mind and spontaneously decide to cross the road after the vehicle. Behave as you would with a real vehicle and driver. Remember to press the button when you want to signal that you want to cross the road. Keep the button pressed as long as you want to cross the road in front of the vehicle.”
	\newpage
	\section{Subjective Results} \label{app2}
	
	\begin{table}[h!]
		\centering
		\begin{tabular}{|c|c|c|c|}
			\hline
			& IAMPDM & RBDM & NIA \\
			\hline
			1  & 8  & 8  & 8  \\
			2  & 1  & 14 & 1  \\
			3  & 11 & 9  & 6  \\
			4  & 11 & 13 & 9  \\
			5  & 9  & 14 & 0  \\
			6  & 10 & 8  & 4  \\
			7  & 7  & 11 & 7  \\
			8  & 6  & 12 & 8  \\
			9  & 14 & 12 & 6  \\
			10 & 9  & 12 & 1  \\
			11 & 7  & 6  & 10 \\
			12 & 14 & 15 & 10 \\
			13 & 12 & 15 & 4  \\
			14 & 9  & 9  & 4  \\
			15 & 10 & 10 & 5  \\
			16 & 11 & 15 & 9  \\
			17 & 9  & 5  & 13 \\
			18 & 10 & 14 & 11 \\
			19 & 10 & 13 & 10 \\
			20 & 4  & 8  & 12 \\
			21 & -& -& -\\
			22 & 11 & 12 & 7  \\
			23 & 15 & 15 & 8  \\
			24 & 13 & 14 & 4  \\
			25 & 1  & 1  & 12 \\
			\hline
		\end{tabular}
		\caption{The subjective assessment for IAMPDM, RBDM, and NIA}
	\end{table}
\end{document}

%% file: res_table_v2crossing.tex
\begin{table}[t!]
\centering
\begin{tabular}{|c|c|c|c|}
\hline
 & $T_{\ml{end}}$ in $s$ & $TTC_{\ml{avg}}$ in $s$ & $DST_{\ml{avg}}$ in $\frac{m}{s^2}$ \\
\hline
\hline
$\mu_{\mathrm{NIA}}$ & 28.49 & 42.09 & 2.02 \\
$\sigma_{\mathrm{ NIA}}$ & 3.47 & 36.75 & 1.33 \\
\hline
\hline
$\mu_{\mathrm{RBDM}}$ & 18.87 & 7.67 & 2.69 \\
$\sigma_{\mathrm{ RBDM}}$ & 5.12 & 0.56 & 1.16 \\
\hline
\hline
$\mu_{\mathrm{IAMPDM}}$ & 27.51 & 7.91 & 3.01 \\
$\sigma_{\mathrm{ IAMPDM}}$ & 17.63 & 1.43 & 0.92 \\
\hline
\hline
\end{tabular}
\caption{Mean values and their standard deviation of the objective metrics in case of \textit{Delayed Crossing}}
\label{tab:res_crossing}
\end{table}

%% file: res_table_v2not_crossing.tex
\begin{table}[t!]
\centering
\begin{tabular}{|c|c|c|c|}
\hline
 & $T_{\ml{end}}$ in $s$ & $TTC_{\ml{avg}}$ in $s$ & $DST_{\ml{avg}}$ in $\frac{m}{s^2}$ \\
\hline
\hline
$\mu_{\mathrm{NIA}}$ & 29.71 & 41.27 & 1.85 \\
$\sigma_{\mathrm{ NIA}}$ & 1.65 & 35.72 & 1.38 \\
\hline
\hline
$\mu_{\mathrm{RBDM}}$ & 16.12 & 7.92 & 3.01 \\
$\sigma_{\mathrm{ RBDM}}$ & 2.56 & 0.17 & 0.57 \\
\hline
\hline
$\mu_{\mathrm{IAMPDM}}$ & 17.18 & 7.59 & 2.71 \\
$\sigma_{\mathrm{ IAMPDM}}$ & 4.09 & 0.10 & 0.37 \\
\hline
\hline
\end{tabular}
\caption{Mean values and their standard deviation of the objective metrics in case of \textit{Delayed Remaining}}
\label{tab:res_remaining}
\end{table}

%% file: varga_IAMPDM_fin_2026_arxiv.bbl
\begin{thebibliography}{54}
\expandafter\ifx\csname natexlab\endcsname\relax\def\natexlab#1{#1}\fi
\providecommand{\url}[1]{\texttt{#1}}
\providecommand{\href}[2]{#2}
\providecommand{\path}[1]{#1}
\providecommand{\DOIprefix}{doi:}
\providecommand{\ArXivprefix}{arXiv:}
\providecommand{\URLprefix}{URL: }
\providecommand{\Pubmedprefix}{pmid:}
\providecommand{\doi}[1]{\href{http://dx.doi.org/#1}{\path{#1}}}
\providecommand{\Pubmed}[1]{\href{pmid:#1}{\path{#1}}}
\providecommand{\bibinfo}[2]{#2}
\ifx\xfnm\relax \def\xfnm[#1]{\unskip,\space#1}\fi
\bibitem[{Amann et~al.(2025)Amann, Probst, Wenzel and
  Weisswange}]{amann2025optimal}
\bibinfo{author}{Amann, M.}, \bibinfo{author}{Probst, M.},
  \bibinfo{author}{Wenzel, R.}, \bibinfo{author}{Weisswange, T.H.},
  \bibinfo{year}{2025}.
\newblock \bibinfo{title}{Optimal behavior planning for implicit communication
  using a probabilistic vehicle–pedestrian interaction model}, in:
  \bibinfo{booktitle}{Proceedings of the IEEE Intelligent Vehicles Symposium
  (IV)}.
\newblock \bibinfo{note}{To appear}.
\bibitem[{Andersson et~al.(2019)Andersson, Gillis, Horn, Rawlings and
  Diehl}]{Andersson2019}
\bibinfo{author}{Andersson, J.A.E.}, \bibinfo{author}{Gillis, J.},
  \bibinfo{author}{Horn, G.}, \bibinfo{author}{Rawlings, J.B.},
  \bibinfo{author}{Diehl, M.}, \bibinfo{year}{2019}.
\newblock \bibinfo{title}{{CasADi} -- {A} software framework for nonlinear
  optimization and optimal control}.
\newblock \bibinfo{journal}{Mathematical Programming Computation}
  \bibinfo{volume}{11}, \bibinfo{pages}{1--36}.
\bibitem[{Azarmi et~al.(2023)Azarmi, Rezaei, Hussain and
  Qian}]{2023_LocalGlobalContextual_azarmi}
\bibinfo{author}{Azarmi, M.}, \bibinfo{author}{Rezaei, M.},
  \bibinfo{author}{Hussain, T.}, \bibinfo{author}{Qian, C.},
  \bibinfo{year}{2023}.
\newblock \bibinfo{title}{Local and {{Global Contextual Features Fusion}} for
  {{Pedestrian Intention Prediction}}}, in: \bibinfo{editor}{Ghatee, M.},
  \bibinfo{editor}{Hashemi, S.M.} (Eds.), \bibinfo{booktitle}{Artificial
  {{Intelligence}} and {{Smart Vehicles}}}. \bibinfo{publisher}{Springer Nature
  Switzerland}, \bibinfo{address}{Cham}. volume \bibinfo{volume}{1883}, pp.
  \bibinfo{pages}{1--13}.
\newblock \DOIprefix\doi{10.1007/978-3-031-43763-2_1}.
\bibitem[{Bindsch{\"a}del et~al.(2021)Bindsch{\"a}del, Krems and
  Kiesel}]{2021_InteractionPedestriansAutomated_bindschadel}
\bibinfo{author}{Bindsch{\"a}del, J.}, \bibinfo{author}{Krems, I.},
  \bibinfo{author}{Kiesel, A.}, \bibinfo{year}{2021}.
\newblock \bibinfo{title}{Interaction between pedestrians and automated
  vehicles: {{Exploring}} a motion-based approach for virtual reality
  experiments}.
\newblock \bibinfo{journal}{Transportation Research Part F: Traffic Psychology
  and Behaviour} \bibinfo{volume}{82}, \bibinfo{pages}{316--332}.
\newblock \DOIprefix\doi{10.1016/j.trf.2021.08.018}.
\bibitem[{Borrelli et~al.(2017)Borrelli, Bemporad and
  Morari}]{2017_PredictiveControlLinear_borrelli}
\bibinfo{author}{Borrelli, F.}, \bibinfo{author}{Bemporad, A.},
  \bibinfo{author}{Morari, M.}, \bibinfo{year}{2017}.
\newblock \bibinfo{title}{Predictive {{Control}} for {{Linear}} and {{Hybrid
  Systems}}:}.
\newblock \bibinfo{edition}{1} ed., \bibinfo{publisher}{Cambridge University
  Press}.
\bibitem[{Bouzidi and Hashemi(2023)}]{2023_IntAware_Merg_Moh}
\bibinfo{author}{Bouzidi, M.K.}, \bibinfo{author}{Hashemi, E.},
  \bibinfo{year}{2023}.
\newblock \bibinfo{title}{Interaction-aware merging in mixed traffic with
  integrated game-theoretic predictive control and inverse differential game},
  in: \bibinfo{booktitle}{2023 IEEE Intelligent Vehicles Symposium (IV)}, pp.
  \bibinfo{pages}{1--6}.
\bibitem[{Camara et~al.(2021)Camara, Bellotto, Cosar, Weber, Nathanael,
  Althoff, Wu, Ruenz, Dietrich, Markkula, Schieben, Tango, Merat and
  Fox}]{2021_PedestrianModelsAutonomous_camara}
\bibinfo{author}{Camara, F.}, \bibinfo{author}{Bellotto, N.},
  \bibinfo{author}{Cosar, S.}, \bibinfo{author}{Weber, F.},
  \bibinfo{author}{Nathanael, D.}, \bibinfo{author}{Althoff, M.},
  \bibinfo{author}{Wu, J.}, \bibinfo{author}{Ruenz, J.},
  \bibinfo{author}{Dietrich, A.}, \bibinfo{author}{Markkula, G.},
  \bibinfo{author}{Schieben, A.}, \bibinfo{author}{Tango, F.},
  \bibinfo{author}{Merat, N.}, \bibinfo{author}{Fox, C.}, \bibinfo{year}{2021}.
\newblock \bibinfo{title}{Pedestrian {{Models}} for {{Autonomous Driving Part
  II}}: {{High-Level Models}} of {{Human Behavior}}}.
\newblock \bibinfo{journal}{IEEE Trans. Intell. Transport. Syst.}
  \bibinfo{volume}{22}, \bibinfo{pages}{5453--5472}.
\newblock \DOIprefix\doi{10.1109/TITS.2020.3006767}.
\bibitem[{Chen et~al.(2023)Chen, Li, Tang, Yang, Cao and
  Lin}]{2023_InteractionAwareDecisionMaking_chen}
\bibinfo{author}{Chen, Y.}, \bibinfo{author}{Li, S.}, \bibinfo{author}{Tang,
  X.}, \bibinfo{author}{Yang, K.}, \bibinfo{author}{Cao, D.},
  \bibinfo{author}{Lin, X.}, \bibinfo{year}{2023}.
\newblock \bibinfo{title}{Interaction-{{Aware Decision Making}} for
  {{Autonomous Vehicles}}}.
\newblock \bibinfo{journal}{IEEE Trans. Transp. Electrific.} ,
  \bibinfo{pages}{1--1}\DOIprefix\doi{10.1109/TTE.2023.3240454}.
\bibitem[{Crosato et~al.(2022)Crosato, Shum, Ho and Wei}]{crosato2022svo}
\bibinfo{author}{Crosato, L.}, \bibinfo{author}{Shum, H.P.H.},
  \bibinfo{author}{Ho, E.S.L.}, \bibinfo{author}{Wei, C.},
  \bibinfo{year}{2022}.
\newblock \bibinfo{title}{Interaction-aware decision-making for automated
  vehicles using social value orientation}.
\newblock \bibinfo{journal}{IEEE Transactions on Intelligent Vehicles}
  \bibinfo{volume}{8}, \bibinfo{pages}{1339--1349}.
\bibitem[{Crosato et~al.(2023)Crosato, Shum, Ho and
  Wei}]{2023_InteractionAwareDecisionMakingAutomated_crosato}
\bibinfo{author}{Crosato, L.}, \bibinfo{author}{Shum, H.P.H.},
  \bibinfo{author}{Ho, E.S.L.}, \bibinfo{author}{Wei, C.},
  \bibinfo{year}{2023}.
\newblock \bibinfo{title}{Interaction-{{Aware Decision-Making}} for {{Automated
  Vehicles Using Social Value Orientation}}}.
\newblock \bibinfo{journal}{IEEE Trans. Intell. Veh.} \bibinfo{volume}{8},
  \bibinfo{pages}{1339--1349}.
\newblock \DOIprefix\doi{10.1109/TIV.2022.3189836}.
\bibitem[{Crosato et~al.(2024)Crosato, Tian, Shum, Ho, Wang and
  Wei}]{crosato2024social}
\bibinfo{author}{Crosato, L.}, \bibinfo{author}{Tian, K.},
  \bibinfo{author}{Shum, H.P.H.}, \bibinfo{author}{Ho, E.S.L.},
  \bibinfo{author}{Wang, Y.}, \bibinfo{author}{Wei, C.}, \bibinfo{year}{2024}.
\newblock \bibinfo{title}{Social interaction-aware dynamical models and
  decision-making for autonomous vehicles}.
\newblock \bibinfo{journal}{Advanced Intelligent Systems} \bibinfo{volume}{6},
  \bibinfo{pages}{2300575}.
\bibitem[{Dalgaard(2008)}]{2008_IntroductoryStatistics_dalgaard}
\bibinfo{author}{Dalgaard, P.}, \bibinfo{year}{2008}.
\newblock \bibinfo{title}{Introductory {{Statistics}} with {{R}}}.
\newblock Statistics and {{Computing}}, \bibinfo{publisher}{Springer New York},
  \bibinfo{address}{New York, NY}.
\bibitem[{Dash et~al.(2023)Dash, Behera, Dehuri and
  Ghosh}]{2023_OutliersDetectionElimination_dash}
\bibinfo{author}{Dash, C.S.K.}, \bibinfo{author}{Behera, A.K.},
  \bibinfo{author}{Dehuri, S.}, \bibinfo{author}{Ghosh, A.},
  \bibinfo{year}{2023}.
\newblock \bibinfo{title}{An outliers detection and elimination framework in
  classification task of data mining}.
\newblock \bibinfo{journal}{Decision Analytics Journal} \bibinfo{volume}{6},
  \bibinfo{pages}{100164}.
\newblock \DOIprefix\doi{10.1016/j.dajour.2023.100164}.
\bibitem[{Deraji{\'c} et~al.(2024)Deraji{\'c}, Bouzidi, Bernhard and
  H{\"o}nig}]{derajic2024learning}
\bibinfo{author}{Deraji{\'c}, B.}, \bibinfo{author}{Bouzidi, M.K.},
  \bibinfo{author}{Bernhard, S.}, \bibinfo{author}{H{\"o}nig, W.},
  \bibinfo{year}{2024}.
\newblock \bibinfo{title}{Learning approximated maximal safe sets via
  hypernetworks for mpc-based local motion planning}.
\newblock \bibinfo{journal}{arXiv preprint arXiv:2410.20267} .
\bibitem[{Di and Shi(2021)}]{2021_SurveyAutonomousVehicle_di}
\bibinfo{author}{Di, X.}, \bibinfo{author}{Shi, R.}, \bibinfo{year}{2021}.
\newblock \bibinfo{title}{A survey on autonomous vehicle control in the era of
  mixed-autonomy: {{From}} physics-based to {{AI-guided}} driving policy
  learning}.
\newblock \bibinfo{journal}{Transportation Research Part C: Emerging
  Technologies} \bibinfo{volume}{125}, \bibinfo{pages}{103008}.
\newblock \DOIprefix\doi{10.1016/j.trc.2021.103008}.
\bibitem[{Domeyer et~al.(2022)Domeyer, Lee, Toyoda, Mehler and
  Reimer}]{2022_DriverPedestrianPerceptualModels_domeyer}
\bibinfo{author}{Domeyer, J.E.}, \bibinfo{author}{Lee, J.D.},
  \bibinfo{author}{Toyoda, H.}, \bibinfo{author}{Mehler, B.},
  \bibinfo{author}{Reimer, B.}, \bibinfo{year}{2022}.
\newblock \bibinfo{title}{Driver-{{Pedestrian Perceptual Models Demonstrate
  Coupling}}: {{Implications}} for {{Vehicle Automation}}}.
\newblock \bibinfo{journal}{IEEE Trans. Human-Mach. Syst.}
  \bibinfo{volume}{52}, \bibinfo{pages}{557--566}.
\newblock \DOIprefix\doi{10.1109/THMS.2022.3158201}.
\bibitem[{{Ezzati Amini} et~al.(2024){Ezzati Amini}, Abouelela, Dhamaniya,
  Friedrich and Antoniou}]{EZZATIAMINI2024107604}
\bibinfo{author}{{Ezzati Amini}, R.}, \bibinfo{author}{Abouelela, M.},
  \bibinfo{author}{Dhamaniya, A.}, \bibinfo{author}{Friedrich, B.},
  \bibinfo{author}{Antoniou, C.}, \bibinfo{year}{2024}.
\newblock \bibinfo{title}{A game-theoretic approach for modelling
  pedestrian–vehicle conflict resolutions in uncontrolled traffic
  environments}.
\newblock \bibinfo{journal}{Accident Analysis \& Prevention}
  \bibinfo{volume}{203}, \bibinfo{pages}{107604}.
\newblock \URLprefix
  \url{https://www.sciencedirect.com/science/article/pii/S0001457524001490},
  \DOIprefix\doi{https://doi.org/10.1016/j.aap.2024.107604}.
\bibitem[{Gulzar et~al.(2021)Gulzar, Muhammad and
  Muhammad}]{2021_SurveyMotionPrediction_gulzar}
\bibinfo{author}{Gulzar, M.}, \bibinfo{author}{Muhammad, Y.},
  \bibinfo{author}{Muhammad, N.}, \bibinfo{year}{2021}.
\newblock \bibinfo{title}{A {{Survey}} on {{Motion Prediction}} of
  {{Pedestrians}} and {{Vehicles}} for {{Autonomous Driving}}}.
\newblock \bibinfo{journal}{IEEE Access} \bibinfo{volume}{9},
  \bibinfo{pages}{137957--137969}.
\newblock \DOIprefix\doi{10.1109/ACCESS.2021.3118224}.
\bibitem[{Hu et~al.(2018)Hu, Zhang and Shelton}]{2018_WhereAreDangerous_hu}
\bibinfo{author}{Hu, Y.}, \bibinfo{author}{Zhang, Y.},
  \bibinfo{author}{Shelton, K.S.}, \bibinfo{year}{2018}.
\newblock \bibinfo{title}{Where are the dangerous intersections for pedestrians
  and cyclists: {{A}} colocation-based approach}.
\newblock \bibinfo{journal}{Transportation Research Part C: Emerging
  Technologies} \bibinfo{volume}{95}, \bibinfo{pages}{431--441}.
\newblock \DOIprefix\doi{10.1016/j.trc.2018.07.030}.
\bibitem[{Kalantari et~al.(2023)Kalantari, Yang, Garcia De~Pedro, Lee,
  Horrobin, Solernou, Holmes, Merat and Markkula}]{2023_WhoGoesFirst_kalantari}
\bibinfo{author}{Kalantari, A.H.}, \bibinfo{author}{Yang, Y.},
  \bibinfo{author}{Garcia De~Pedro, J.}, \bibinfo{author}{Lee, Y.M.},
  \bibinfo{author}{Horrobin, A.}, \bibinfo{author}{Solernou, A.},
  \bibinfo{author}{Holmes, C.}, \bibinfo{author}{Merat, N.},
  \bibinfo{author}{Markkula, G.}, \bibinfo{year}{2023}.
\newblock \bibinfo{title}{Who goes first? {{A}} distributed simulator study of
  vehicle--pedestrian interaction}.
\newblock \bibinfo{journal}{Accident Analysis \& Prevention}
  \bibinfo{volume}{186}, \bibinfo{pages}{107050}.
\newblock \DOIprefix\doi{10.1016/j.aap.2023.107050}.
\bibitem[{Kampitakis et~al.(2023)Kampitakis, Fafoutellis, Oprea and
  Vlahogianni}]{2023_SharedSpaceMultimodal_kampitakis}
\bibinfo{author}{Kampitakis, E.P.}, \bibinfo{author}{Fafoutellis, P.},
  \bibinfo{author}{Oprea, G.M.}, \bibinfo{author}{Vlahogianni, E.I.},
  \bibinfo{year}{2023}.
\newblock \bibinfo{title}{Shared space multi-modal traffic modeling using
  {{LSTM}} networks with repulsion map and an intention-based multi-loss
  function}.
\newblock \bibinfo{journal}{Transportation Research Part C: Emerging
  Technologies} \bibinfo{volume}{150}, \bibinfo{pages}{104104}.
\newblock \DOIprefix\doi{10.1016/j.trc.2023.104104}.
\bibitem[{Korbmacher and
  Tordeux(2022)}]{2022_ReviewPedestrianTrajectory_korbmacher}
\bibinfo{author}{Korbmacher, R.}, \bibinfo{author}{Tordeux, A.},
  \bibinfo{year}{2022}.
\newblock \bibinfo{title}{Review of {{Pedestrian Trajectory Prediction
  Methods}}: {{Comparing Deep Learning}} and {{Knowledge-Based Approaches}}}.
\newblock \bibinfo{journal}{IEEE Trans. Intell. Transport. Syst.}
  \bibinfo{volume}{23}, \bibinfo{pages}{24126--24144}.
\newblock \DOIprefix\doi{10.1109/TITS.2022.3205676}.
\bibitem[{L{\"o}cken et~al.(2019)L{\"o}cken, Golling and
  Riener}]{2019_HowShouldAutomated_locken}
\bibinfo{author}{L{\"o}cken, A.}, \bibinfo{author}{Golling, C.},
  \bibinfo{author}{Riener, A.}, \bibinfo{year}{2019}.
\newblock \bibinfo{title}{How {{Should Automated Vehicles Interact}} with
  {{Pedestrians}}?: {{A Comparative Analysis}} of {{Interaction Concepts}} in
  {{Virtual Reality}}}, in: \bibinfo{booktitle}{Proceedings of the 11th
  {{International Conference}} on {{Automotive User Interfaces}} and
  {{Interactive Vehicular Applications}}}, \bibinfo{publisher}{ACM},
  \bibinfo{address}{Utrecht Netherlands}. pp. \bibinfo{pages}{262--274}.
\newblock \DOIprefix\doi{10.1145/3342197.3344544}.
\bibitem[{Luu et~al.(2022)Luu, Eom, Cho, Kim, Oh and
  Kim}]{2022_CautiousBehaviorsPedestrians_luu}
\bibinfo{author}{Luu, D.T.}, \bibinfo{author}{Eom, H.}, \bibinfo{author}{Cho,
  G.H.}, \bibinfo{author}{Kim, S.N.}, \bibinfo{author}{Oh, J.},
  \bibinfo{author}{Kim, J.}, \bibinfo{year}{2022}.
\newblock \bibinfo{title}{Cautious behaviors of pedestrians while crossing
  narrow streets: {{Exploration}} of behaviors using virtual reality
  experiments}.
\newblock \bibinfo{journal}{Transportation Research Part F: Traffic Psychology
  and Behaviour} \bibinfo{volume}{91}, \bibinfo{pages}{164--178}.
\newblock \DOIprefix\doi{10.1016/j.trf.2022.09.024}.
\bibitem[{{Mar{\'i}n-Solano}(2021)}]{2021_DynamicBargainingTimeConsistency_marin-solano}
\bibinfo{author}{{Mar{\'i}n-Solano}, J.}, \bibinfo{year}{2021}.
\newblock \bibinfo{title}{Dynamic {{Bargaining}} and {{Time-Consistency}} in
  {{Linear-State}} and {{Homogeneous Linear-Quadratic Cooperative Differential
  Games}}}.
\newblock \bibinfo{journal}{Int. Game Theory Rev.} \bibinfo{volume}{23},
  \bibinfo{pages}{2250003}.
\newblock \DOIprefix\doi{10.1142/S0219198922500037}.
\bibitem[{Markkula et~al.(2020a)Markkula, Madigan, Nathanael, Portouli, Lee,
  Dietrich, Billington, Schieben and Merat}]{markkula2020defining}
\bibinfo{author}{Markkula, G.}, \bibinfo{author}{Madigan, R.},
  \bibinfo{author}{Nathanael, D.}, \bibinfo{author}{Portouli, E.},
  \bibinfo{author}{Lee, Y.M.}, \bibinfo{author}{Dietrich, A.},
  \bibinfo{author}{Billington, J.}, \bibinfo{author}{Schieben, A.},
  \bibinfo{author}{Merat, N.}, \bibinfo{year}{2020}a.
\newblock \bibinfo{title}{Defining interactions: a conceptual framework for
  understanding interactive behaviour in human and automated road traffic}.
\newblock \bibinfo{journal}{Theoretical Issues in Ergonomics Science}
  \bibinfo{volume}{21}, \bibinfo{pages}{728--752}.
\bibitem[{Markkula et~al.(2020b)Markkula, Madigan, Nathanael, Portouli, Lee,
  Dietrich, Billington, Schieben and
  Merat}]{2020_DefiningInteractionsConceptual_markkula}
\bibinfo{author}{Markkula, G.}, \bibinfo{author}{Madigan, R.},
  \bibinfo{author}{Nathanael, D.}, \bibinfo{author}{Portouli, E.},
  \bibinfo{author}{Lee, Y.M.}, \bibinfo{author}{Dietrich, A.},
  \bibinfo{author}{Billington, J.}, \bibinfo{author}{Schieben, A.},
  \bibinfo{author}{Merat, N.}, \bibinfo{year}{2020}b.
\newblock \bibinfo{title}{Defining interactions: A conceptual framework for
  understanding interactive behaviour in human and automated road traffic}.
\newblock \bibinfo{journal}{Theoretical Issues in Ergonomics Science}
  \bibinfo{volume}{21}, \bibinfo{pages}{728--752}.
\newblock \DOIprefix\doi{10.1080/1463922X.2020.1736686}.
\bibitem[{Pavelko et~al.(2022)Pavelko, Skugor, Deur, Ivanovic and
  Tseng}]{2022_GameTheoryBasedModeling_pavelko}
\bibinfo{author}{Pavelko, L.}, \bibinfo{author}{Skugor, B.},
  \bibinfo{author}{Deur, J.}, \bibinfo{author}{Ivanovic, V.},
  \bibinfo{author}{Tseng, H.E.}, \bibinfo{year}{2022}.
\newblock \bibinfo{title}{Game {{Theory-Based Modeling}} of
  {{Multi-Vehicle}}/{{Multi-Pedestrian Interaction}} at {{Unsignalized
  Crosswalks}}}.
\newblock \bibinfo{journal}{SAE Int. J. Adv. \& Curr. Prac. in Mobility}
  \bibinfo{volume}{4}, \bibinfo{pages}{2469--2477}.
\newblock \DOIprefix\doi{10.4271/2022-01-0814}.
\bibitem[{Predhumeau et~al.(2023)Predhumeau, Spalanzani and
  Dugdale}]{2023_PedestrianBehaviorShared_predhumeau}
\bibinfo{author}{Predhumeau, M.}, \bibinfo{author}{Spalanzani, A.},
  \bibinfo{author}{Dugdale, J.}, \bibinfo{year}{2023}.
\newblock \bibinfo{title}{Pedestrian {{Behavior}} in {{Shared Spaces With
  Autonomous Vehicles}}: {{An Integrated Framework}} and {{Review}}}.
\newblock \bibinfo{journal}{IEEE Trans. Intell. Veh.} \bibinfo{volume}{8},
  \bibinfo{pages}{438--457}.
\newblock \DOIprefix\doi{10.1109/TIV.2021.3116436}.
\bibitem[{{Raghuram Kadali} et~al.(2014){Raghuram Kadali}, Rathi and
  Perumal}]{2014_PedMidBlock_JTTE}
\bibinfo{author}{{Raghuram Kadali}, B.}, \bibinfo{author}{Rathi, N.},
  \bibinfo{author}{Perumal, V.}, \bibinfo{year}{2014}.
\newblock \bibinfo{title}{Evaluation of pedestrian mid-block road crossing
  behaviour using artificial neural network}.
\newblock \bibinfo{journal}{Journal of Traffic and Transportation Engineering
  (English Edition)} \bibinfo{volume}{1}, \bibinfo{pages}{111--119}.
\bibitem[{Rasouli et~al.(2022)Rasouli, Yau, Rohani and
  Luo}]{2022_MultiModalHybridArchitecture_rasouli}
\bibinfo{author}{Rasouli, A.}, \bibinfo{author}{Yau, T.},
  \bibinfo{author}{Rohani, M.}, \bibinfo{author}{Luo, J.},
  \bibinfo{year}{2022}.
\newblock \bibinfo{title}{Multi-{{Modal Hybrid Architecture}} for {{Pedestrian
  Action Prediction}}}, in: \bibinfo{booktitle}{2022 {{IEEE Intelligent
  Vehicles Symposium}} ({{IV}})}, \bibinfo{publisher}{IEEE},
  \bibinfo{address}{Aachen, Germany}. pp. \bibinfo{pages}{91--97}.
\newblock \DOIprefix\doi{10.1109/IV51971.2022.9827055}.
\bibitem[{Razali et~al.(2021)Razali, Mordan and
  Alahi}]{2021_PedestrianIntentionPrediction_razali}
\bibinfo{author}{Razali, H.}, \bibinfo{author}{Mordan, T.},
  \bibinfo{author}{Alahi, A.}, \bibinfo{year}{2021}.
\newblock \bibinfo{title}{Pedestrian intention prediction: {{A}} convolutional
  bottom-up multi-task approach}.
\newblock \bibinfo{journal}{Transportation Research Part C: Emerging
  Technologies} \bibinfo{volume}{130}, \bibinfo{pages}{103259}.
\newblock \DOIprefix\doi{10.1016/j.trc.2021.103259}.
\bibitem[{Razmi~Rad et~al.(2020)Razmi~Rad, Homem De Almeida~Correia and
  Hagenzieker}]{2020_PedestriansRoadCrossing_razmirad}
\bibinfo{author}{Razmi~Rad, S.}, \bibinfo{author}{Homem De Almeida~Correia,
  G.}, \bibinfo{author}{Hagenzieker, M.}, \bibinfo{year}{2020}.
\newblock \bibinfo{title}{Pedestrians' road crossing behaviour in front of
  automated vehicles: {{Results}} from a pedestrian simulation experiment using
  agent-based modelling}.
\newblock \bibinfo{journal}{Transportation Research Part F: Traffic Psychology
  and Behaviour} \bibinfo{volume}{69}, \bibinfo{pages}{101--119}.
\newblock \DOIprefix\doi{10.1016/j.trf.2020.01.014}.
\bibitem[{Rezaee et~al.(2021)Rezaee, Yadmellat and
  Chamorro}]{2021_MotionPlanningAutonomous_rezaee}
\bibinfo{author}{Rezaee, K.}, \bibinfo{author}{Yadmellat, P.},
  \bibinfo{author}{Chamorro, S.}, \bibinfo{year}{2021}.
\newblock \bibinfo{title}{Motion {{Planning}} for {{Autonomous Vehicles}} in
  the {{Presence}} of {{Uncertainty Using Reinforcement Learning}}}, in:
  \bibinfo{booktitle}{2021 {{IEEE}}/{{RSJ International Conference}} on
  {{Intelligent Robots}} and {{Systems}} ({{IROS}})},
  \bibinfo{publisher}{IEEE}, \bibinfo{address}{Prague, Czech Republic}. pp.
  \bibinfo{pages}{3506--3511}.
\newblock \DOIprefix\doi{10.1109/IROS51168.2021.9636480}.
\bibitem[{Russo et~al.(2021)Russo, Terlizzi, Tipaldi and
  Glielmo}]{2021_ReinforcementLearningApproach_russo}
\bibinfo{author}{Russo, L.}, \bibinfo{author}{Terlizzi, M.},
  \bibinfo{author}{Tipaldi, M.}, \bibinfo{author}{Glielmo, L.},
  \bibinfo{year}{2021}.
\newblock \bibinfo{title}{A {{Reinforcement Learning}} approach for pedestrian
  collision avoidance and trajectory tracking in autonomous driving systems},
  in: \bibinfo{booktitle}{2021 5th {{International Conference}} on {{Control}}
  and {{Fault-Tolerant Systems}} ({{SysTol}})}, \bibinfo{publisher}{IEEE},
  \bibinfo{address}{Saint-Raphael, France}. pp. \bibinfo{pages}{44--49}.
\newblock \DOIprefix\doi{10.1109/SysTol52990.2021.9595150}.
\bibitem[{{\v S}kugor et~al.(2023){\v S}kugor, Deur, Ivanovic and
  Tseng}]{2023_StochasticModelPredictive_skugor}
\bibinfo{author}{{\v S}kugor, B.}, \bibinfo{author}{Deur, J.},
  \bibinfo{author}{Ivanovic, V.}, \bibinfo{author}{Tseng, H.E.},
  \bibinfo{year}{2023}.
\newblock \bibinfo{title}{Stochastic {{Model Predictive Control}} of an
  {{Autonomous Vehicle Interacting}} with {{Pedestrians}} at {{Unsignalized
  Crosswalks}}}, in: \bibinfo{booktitle}{2023 {{European Control Conference}}
  ({{ECC}})}, \bibinfo{publisher}{IEEE}, \bibinfo{address}{Bucharest, Romania}.
  pp. \bibinfo{pages}{1--7}.
\newblock \DOIprefix\doi{10.23919/ECC57647.2023.10178144}.
\bibitem[{Skugor et~al.(2020)Skugor, Topic, Deur, Ivanovic and
  Tseng}]{2020_AnalysisGameTheorybased_skugor}
\bibinfo{author}{Skugor, B.}, \bibinfo{author}{Topic, J.},
  \bibinfo{author}{Deur, J.}, \bibinfo{author}{Ivanovic, V.},
  \bibinfo{author}{Tseng, E.}, \bibinfo{year}{2020}.
\newblock \bibinfo{title}{Analysis of a {{Game Theory-based Model}} of
  {{Vehicle-Pedestrian Interaction}} at {{Uncontrolled Crosswalks}}}, in:
  \bibinfo{booktitle}{2020 {{International Conference}} on {{Smart Systems}}
  and {{Technologies}} ({{SST}})}, \bibinfo{publisher}{IEEE},
  \bibinfo{address}{Osijek, Croatia}. pp. \bibinfo{pages}{73--81}.
\newblock \DOIprefix\doi{10.1109/SST49455.2020.9264131}.
\bibitem[{Tian et~al.(2024)Tian, Markkula, Wei, Lee, Madigan, Hirose, Merat and
  Romano}]{2024_DconstrHuman_tian}
\bibinfo{author}{Tian, K.}, \bibinfo{author}{Markkula, G.},
  \bibinfo{author}{Wei, C.}, \bibinfo{author}{Lee, Y.M.},
  \bibinfo{author}{Madigan, R.}, \bibinfo{author}{Hirose, T.},
  \bibinfo{author}{Merat, N.}, \bibinfo{author}{Romano, R.},
  \bibinfo{year}{2024}.
\newblock \bibinfo{title}{Deconstructing pedestrian crossing decisions in
  interactions with continuous traffic: An anthropomorphic model}.
\newblock \bibinfo{journal}{IEEE Transactions on Intelligent Transportation
  Systems} \bibinfo{volume}{25}, \bibinfo{pages}{2466--2478}.
\bibitem[{Tian et~al.(2023)Tian, Tzigieras, Wei, Lee, Holmes, Leonetti, Merat,
  Romano and Markkula}]{tian2023deceleration}
\bibinfo{author}{Tian, K.}, \bibinfo{author}{Tzigieras, A.},
  \bibinfo{author}{Wei, C.}, \bibinfo{author}{Lee, Y.M.},
  \bibinfo{author}{Holmes, C.}, \bibinfo{author}{Leonetti, M.},
  \bibinfo{author}{Merat, N.}, \bibinfo{author}{Romano, R.},
  \bibinfo{author}{Markkula, G.}, \bibinfo{year}{2023}.
\newblock \bibinfo{title}{Deceleration parameters as implicit communication
  signals for pedestrians' crossing decisions and estimations of automated
  vehicle behaviour}.
\newblock \bibinfo{journal}{Accident Analysis and Prevention}
  \bibinfo{volume}{190}, \bibinfo{pages}{107173}.
\bibitem[{Tian et~al.(2025a)Tian, Tzigieras, Wei, Lee, Holmes, Leonetti, Merat,
  Romano and Markkula}]{tian2025interacting}
\bibinfo{author}{Tian, K.}, \bibinfo{author}{Tzigieras, A.},
  \bibinfo{author}{Wei, C.}, \bibinfo{author}{Lee, Y.M.},
  \bibinfo{author}{Holmes, C.}, \bibinfo{author}{Leonetti, M.},
  \bibinfo{author}{Merat, N.}, \bibinfo{author}{Romano, R.},
  \bibinfo{author}{Markkula, G.}, \bibinfo{year}{2025}a.
\newblock \bibinfo{title}{Interacting with yielding vehicles: a perceptually
  plausible model for pedestrian road crossing decisions}.
\newblock \bibinfo{journal}{IEEE Transactions on Intelligent Transportation
  Systems} \bibinfo{note}{Accepted for publication}.
\bibitem[{Tian et~al.(2025b)Tian, Wu, Qiu, Wu et~al.}]{tian2025tradeoff}
\bibinfo{author}{Tian, K.}, \bibinfo{author}{Wu, J.}, \bibinfo{author}{Qiu,
  T.Z.}, \bibinfo{author}{Wu, C.}, et~al., \bibinfo{year}{2025}b.
\newblock \bibinfo{title}{A framework for analyzing driver safety-efficiency
  trade-offs at uncontrolled crosswalks: Towards social vehicle automation}.
\newblock \bibinfo{journal}{Safety Science} \bibinfo{volume}{187},
  \bibinfo{pages}{106860}.
\bibitem[{Tran et~al.(2021)Tran, Parker and
  Tomitsch}]{2021_ReviewVirtualReality_tran}
\bibinfo{author}{Tran, T.}, \bibinfo{author}{Parker, C.},
  \bibinfo{author}{Tomitsch, M.}, \bibinfo{year}{2021}.
\newblock \bibinfo{title}{A {{Review}} of {{Virtual Reality Studies}} on
  {{Autonomous Vehicle}}--{{Pedestrian Interaction}}}.
\newblock \bibinfo{journal}{IEEE Trans. Human-Mach. Syst.}
  \bibinfo{volume}{51}, \bibinfo{pages}{641--652}.
\newblock \DOIprefix\doi{10.1109/THMS.2021.3107517}.
\bibitem[{Trumpp et~al.(2022)Trumpp, Bayerlein and
  Gesbert}]{2022_ModelingInteractionsAutonomous_trumpp}
\bibinfo{author}{Trumpp, R.}, \bibinfo{author}{Bayerlein, H.},
  \bibinfo{author}{Gesbert, D.}, \bibinfo{year}{2022}.
\newblock \bibinfo{title}{Modeling {{Interactions}} of {{Autonomous Vehicles}}
  and {{Pedestrians}} with {{Deep Multi-Agent Reinforcement Learning}} for
  {{Collision Avoidance}}}, in: \bibinfo{booktitle}{2022 {{IEEE Intelligent
  Vehicles Symposium}} ({{IV}})}, \bibinfo{publisher}{IEEE},
  \bibinfo{address}{Aachen, Germany}. pp. \bibinfo{pages}{331--336}.
\newblock \DOIprefix\doi{10.1109/IV51971.2022.9827451}.
\bibitem[{Varga(2024)}]{2024_AdaptiveCooperationModelBased_varga}
\bibinfo{author}{Varga, B.}, \bibinfo{year}{2024}.
\newblock \bibinfo{title}{Toward {{Adaptive Cooperation}}: {{Model-Based Shared
  Control Using LQ-Differential Games}}}.
\newblock \bibinfo{journal}{ACTA POLYTECH HUNG} \bibinfo{volume}{21},
  \bibinfo{pages}{439--456}.
\newblock \DOIprefix\doi{10.12700/APH.21.10.2024.10.27}.
\bibitem[{Varga et~al.(2023a)Varga, Yang and
  Hohmann}]{2023_IntentionAwareDecisionMakingMixed_varga}
\bibinfo{author}{Varga, B.}, \bibinfo{author}{Yang, D.},
  \bibinfo{author}{Hohmann, S.}, \bibinfo{year}{2023}a.
\newblock \bibinfo{title}{Intention-{{Aware Decision-Making}} for {{Mixed
  Intersection Scenarios}}}, in: \bibinfo{booktitle}{2023 {{IEEE}} 17th
  {{International Symposium}} on {{Applied Computational Intelligence}} and
  {{Informatics}} ({{SACI}})}, \bibinfo{publisher}{IEEE},
  \bibinfo{address}{Timisoara, Romania}. pp. \bibinfo{pages}{000369--000374}.
\newblock \DOIprefix\doi{10.1109/SACI58269.2023.10158550}.
\bibitem[{Varga et~al.(2023b)Varga, Yang, Martin and
  Hohmann}]{2023_CooperativeDecisionMakingShared_varga}
\bibinfo{author}{Varga, B.}, \bibinfo{author}{Yang, D.},
  \bibinfo{author}{Martin, M.}, \bibinfo{author}{Hohmann, S.},
  \bibinfo{year}{2023}b.
\newblock \bibinfo{title}{Cooperative {{Decision-Making}} in {{Shared Spaces}}:
  {{Making Urban Traffic Safer Through Human-Machine Cooperation}}}, in:
  \bibinfo{booktitle}{2023 {{IEEE}} 21st {{Jubilee International Symposium}} on
  {{Intelligent Systems}} and {{Informatics}} ({{SISY}})},
  \bibinfo{publisher}{IEEE}, \bibinfo{address}{Pula, Croatia}. pp.
  \bibinfo{pages}{000109--000114}.
\newblock \DOIprefix\doi{10.1109/SISY60376.2023.10417908}.
\bibitem[{Wang et~al.(2022)Wang, Rong, Sun and Liu}]{wang2022social}
\bibinfo{author}{Wang, W.}, \bibinfo{author}{Rong, L.}, \bibinfo{author}{Sun,
  S.}, \bibinfo{author}{Liu, D.}, \bibinfo{year}{2022}.
\newblock \bibinfo{title}{Social interactions for autonomous driving: a review
  and perspectives}.
\newblock \bibinfo{journal}{Foundations and Trends in Robotics}
  \bibinfo{volume}{10}, \bibinfo{pages}{198--376}.
\bibitem[{Westhofen et~al.(2023)Westhofen, Neurohr, Koopmann, Butz, Sch{\"u}tt,
  Utesch, Neurohr, Gutenkunst and
  B{\"o}de}]{2023_CriticalityMetricsAutomated_westhofen}
\bibinfo{author}{Westhofen, L.}, \bibinfo{author}{Neurohr, C.},
  \bibinfo{author}{Koopmann, T.}, \bibinfo{author}{Butz, M.},
  \bibinfo{author}{Sch{\"u}tt, B.}, \bibinfo{author}{Utesch, F.},
  \bibinfo{author}{Neurohr, B.}, \bibinfo{author}{Gutenkunst, C.},
  \bibinfo{author}{B{\"o}de, E.}, \bibinfo{year}{2023}.
\newblock \bibinfo{title}{Criticality {{Metrics}} for {{Automated Driving}}:
  {{A Review}} and {{Suitability Analysis}} of the {{State}} of the {{Art}}}.
\newblock \bibinfo{journal}{Arch Computat Methods Eng} \bibinfo{volume}{30},
  \bibinfo{pages}{1--35}.
\newblock \DOIprefix\doi{10.1007/s11831-022-09788-7}.
\bibitem[{Yang et~al.(2022)Yang, Zhan, Wang, Chan, Cai and
  Wang}]{2022_CrossingNotContextBased_yang}
\bibinfo{author}{Yang, B.}, \bibinfo{author}{Zhan, W.}, \bibinfo{author}{Wang,
  P.}, \bibinfo{author}{Chan, C.}, \bibinfo{author}{Cai, Y.},
  \bibinfo{author}{Wang, N.}, \bibinfo{year}{2022}.
\newblock \bibinfo{title}{Crossing or {{Not}}? {{Context-Based Recognition}} of
  {{Pedestrian Crossing Intention}} in the {{Urban Environment}}}.
\newblock \bibinfo{journal}{IEEE Trans. Intell. Transport. Syst.}
  \bibinfo{volume}{23}, \bibinfo{pages}{5338--5349}.
\newblock \DOIprefix\doi{10.1109/TITS.2021.3053031}.
\bibitem[{Yang et~al.(2019)Yang, Li, Redmill and Özgüner}]{2019_IV_Yang}
\bibinfo{author}{Yang, D.}, \bibinfo{author}{Li, L.}, \bibinfo{author}{Redmill,
  K.}, \bibinfo{author}{Özgüner, m.}, \bibinfo{year}{2019}.
\newblock \bibinfo{title}{Top-view trajectories: A pedestrian dataset of
  vehicle-crowd interaction from controlled experiments and crowded campus},
  in: \bibinfo{booktitle}{2019 IEEE Intelligent Vehicles Symposium (IV)}, pp.
  \bibinfo{pages}{899--904}.
\bibitem[{Yang et~al.(2024)Yang, Lee, Madigan, Solernou and
  Merat}]{2024_InterpretingPedestriansHead_yang}
\bibinfo{author}{Yang, Y.}, \bibinfo{author}{Lee, Y.M.},
  \bibinfo{author}{Madigan, R.}, \bibinfo{author}{Solernou, A.},
  \bibinfo{author}{Merat, N.}, \bibinfo{year}{2024}.
\newblock \bibinfo{title}{Interpreting pedestrians' head movements when
  encountering automated vehicles at a virtual crossroad}.
\newblock \bibinfo{journal}{Transportation Research Part F: Traffic Psychology
  and Behaviour} \bibinfo{volume}{103}, \bibinfo{pages}{340--352}.
\newblock \DOIprefix\doi{10.1016/j.trf.2024.04.022}.
\bibitem[{Zhang et~al.(2022)Zhang, Ma, Wang, Hinz and
  Knoll}]{2022_EfficientPOMDPBehavior_zhang}
\bibinfo{author}{Zhang, C.}, \bibinfo{author}{Ma, S.}, \bibinfo{author}{Wang,
  M.}, \bibinfo{author}{Hinz, G.}, \bibinfo{author}{Knoll, A.},
  \bibinfo{year}{2022}.
\newblock \bibinfo{title}{Efficient {{POMDP Behavior Planning}} for
  {{Autonomous Driving}} in {{Dense Urban Environments}} using {{Multi-Step
  Occupancy Grid Maps}}}, in: \bibinfo{booktitle}{2022 {{IEEE}} 25th
  {{International Conference}} on {{Intelligent Transportation Systems}}
  ({{ITSC}})}, \bibinfo{publisher}{IEEE}, \bibinfo{address}{Macau, China}. pp.
  \bibinfo{pages}{2722--2729}.
\newblock \DOIprefix\doi{10.1109/ITSC55140.2022.9922353}.
\bibitem[{Zhou et~al.(2022)Zhou, Sun, Liu and
  Burnett}]{2022_FactorsAffectingPedestrians_zhou}
\bibinfo{author}{Zhou, S.}, \bibinfo{author}{Sun, X.}, \bibinfo{author}{Liu,
  B.}, \bibinfo{author}{Burnett, G.}, \bibinfo{year}{2022}.
\newblock \bibinfo{title}{Factors {{Affecting Pedestrians}}' {{Trust}} in
  {{Automated Vehicles}}: {{Literature Review}} and {{Theoretical Model}}}.
\newblock \bibinfo{journal}{IEEE Trans. Human-Mach. Syst.}
  \bibinfo{volume}{52}, \bibinfo{pages}{490--500}.
\newblock \DOIprefix\doi{10.1109/THMS.2021.3112956}.
\bibitem[{Zhou et~al.(2024)Zhou, Liu, Liu, Ouyang and
  Tang}]{2024_PedestrianCrossingIntention_zhou}
\bibinfo{author}{Zhou, Z.}, \bibinfo{author}{Liu, Y.}, \bibinfo{author}{Liu,
  B.}, \bibinfo{author}{Ouyang, M.}, \bibinfo{author}{Tang, R.},
  \bibinfo{year}{2024}.
\newblock \bibinfo{title}{Pedestrian {{Crossing Intention Prediction Model
  Considering Social Interaction}} between {{Multi-Pedestrians}} and
  {{Multi-Vehicles}}}.
\newblock \bibinfo{journal}{Transportation Research Record: Journal of the
  Transportation Research Board} \bibinfo{volume}{2678},
  \bibinfo{pages}{80--101}.

\end{thebibliography}
